\documentclass[12pt]{article}
 \usepackage{secdot}
\usepackage{graphicx}
\usepackage{epstopdf}
\usepackage{setspace}
\usepackage{amssymb,amsmath,array,tabularx,verbatim,dsfont}
\usepackage[latin1]{inputenc}
\usepackage[T1]{fontenc}
\usepackage{bm}
\usepackage[authoryear]{natbib}
\usepackage{color}
\usepackage{subfigure}
\usepackage{comment}

\usepackage{setspace}
\usepackage{xcolor}

\newcommand{\bea}{\begin{eqnarray*}}
\newcommand{\eea}{\end{eqnarray*}}

\newcommand{\tr}{\mathrm{tr}}
\newcommand{\R}{\mathbb{R}}
\newcommand{\D}{\mathcal{D}}

\newcommand{\I}{\mathcal{I}}
\newcommand{\Cov}{\mathrm{Cov}}

\newtheorem{Theorem}{\bf Theorem}[section]

\newtheorem{Remark}{\bf Remark}[section]

\newtheorem{Example}{Example}[section]
\newtheorem{Proposition}{Proposition}[section]

\begin{document}

\parindent 0cm
\title{Optimal designs for active controlled dose finding trials  with   efficacy-toxicity outcomes  }

\author{
{\small Holger Dette, Katrin Kettelhake, Kirsten Schorning } \\
{\small Ruhr-Universit\"at Bochum } \\
{\small Fakult\"at f\"ur Mathematik } \\
{\small 44780 Bochum, Germany } \\
{\small e-mail: holger.dette@ruhr-uni-bochum.de }\\
\and
{\small Weng Kee Wong } \\
{\small Department of Biostatistics } \\
{\small UCLA School of Public Health } \\
{\small  Los Angeles, CA 90095-1772 } \\
{\small e-mail: wkwong@ucla.edu }\\
\and
{\small Frank Bretz } \\
{\small Statistical Methodology } \\
{\small  Novartis Pharma AG } \\
{\small 4002 Basel, Switzerland } \\
{\small e-mail: frank.bretz@novartis.com}
}

\maketitle
\begin{abstract}
Nonlinear regression models addressing both efficacy and toxicity outcomes are increasingly used in dose-finding trials, such as in pharmaceutical drug development. However, research on related experimental design problems for corresponding active controlled trials is still scarce. In this paper we derive optimal designs to estimate efficacy and toxicity in an active controlled clinical dose finding 
trial when the bivariate continuous outcomes are modeled either by polynomials up to degree 2, the Michaelis-Menten model, the Emax model, or a combination thereof.  We determine  upper bounds on the number of different doses levels required
for the optimal design and provide conditions under which the boundary points of the design space are included in the optimal design.
We also provide an analytical description of the minimally supported D-optimal designs and show that they do not depend
 on the correlation between the bivariate outcomes. 
We illustrate the  proposed methods with numerical examples and demonstrate the advantages of the $D$-optimal design  
for a trial, which has recently been considered in the literature.
\end{abstract}


% \doublespacing

%%%%%%%%%------------------------------------%%%%%%%%%%%%%%%%%%%%%%%%%%%%%%%%

\section{Introduction}
\def\theequation{1.\arabic{equation}}
\setcounter{equation}{0}
There is a vast literature on optimal design of experiments, with applications ranging across many disciplines. \cite{atkinson1996} showed the usefulness of optimal designs for real applications using examples from agriculture, animal breeding studies, accelerated life-testing experiments and computer experiments.  \cite{berger2005} gave examples of the varied disciplines that increasingly use optimal design ideas for scientific investigations.  Most of the literature focuses on optimal designs for models with a univariate outcome. In practice, however, drug trials are often conducted to measure multiple outcomes that are likely to be correlated.
%studies have several outcomes and usually they are correlated. For instance, in a randomized clinical trial for scleroderma patients, multiple laboratory measurements on lung function, %inflammation markers and clinical  outcomes such as Rodnam skin score, health assessment questionnaire, ulcers and flares are observed longitudinally.  For controlled trials, such as early %phase trials, statistical models increasingly incorporate efficacy and toxicity outcomes in the design of the trial. An increasingly  typical design goal for an early phase trial is to %determine the optimal number and the right dose levels to observe the correlated bivarate  outcomes and make the most accurate inference on the model parameters or interesting functions of %the model parameters.
For instance, pharmaceutical dose-finding trials invariably have bivariate outcomes involving efficacy and toxicity. As a motivating example consider a
randomized controlled clinical trials for hypertensive patients treated with an angiotensin-converting-enzyme (ACE) inhibitor. In such trials change from baseline in the sitting blood pressure is a frequent efficacy outcome. However, patients starting on an ACE inhibitor usually have a modest reduction in glomerular filtration rate (GFR) that stabilizes after several days. Because this decrease may be significant in conditions of decreased renal perfusion, the renal function should be closely monitored over the first few days in those patients [see \cite{sidnav2014} and \cite{talipi2015}].
%\footnote{New reference: Sidorenkov G, Navis G (2014). Safety of ACE inhibitor therapies in patients with chronic kidney disease. Expert Opinion on Drug Safety 13 (10): 1383-1395}
Thus, the amount of decrease in GFR from baseline is a common outcome to assess unwanted side effects.

Several papers have addressed design problems that incorporate both efficacy and toxicity of a drug.
\cite{fancha2004} proposed using a continuous ratio model for a trinomial outcome, where the outcome of a patient may be classified as ``no reaction'' when neither toxicity nor efficacy occurs, ``efficacy'' for efficacy without toxicity, and ``adverse reaction'' for toxicity. \cite{heisemyers1996} used the Gumbel bivariate binary quantal response model to study efficacy and toxicity.  In their example, patients were continuously monitored whether they experienced a toxicity event or a treatment benefit. The authors investigated locally D- and Q-optimal designs, where D-optimal designs were constructed  for estimating all model parameters and Q-optimal designs were obtained by maximizing the probability of efficacy without toxicity at a selected dose level. More recently, \cite{ireland2013}  applied c-optimal designs to the bivariate Emax model for continuous efficacy and toxicity outcomes. The author  determined the dose level providing the best possible combination of efficacy and toxicity, based on a pre-specified clinical utility index [see \cite{carrothers2011}].
%The CUI is somewhat similar to a desirability function, which is a goodness  measure for capturing the multiple dimensions of a product commonly used in the manufacturing sector.  CUI is %more geared toward pharmaceutical and clinical applications and  is a practical and useful approach for informing drug development decisions where multiple aspects of a drug's clinical %profile must be optimally balanced.

Adaptive dose finding trials incorporating both efficacy and safety have been investigated as well.
For example, \cite{thall2004} and \cite{drafed2006} found adaptive designs for dose-finding based on efficacy-toxicity outcome using a Gumbel bivariate logistic regression or a Cox bivariate model. \cite{drafewu2008} proposed new designs for selecting drug combinations for a bivariate Probit correlated model based on an efficacy-toxicity outcome profile of a drug using Bayesian, minimax and adaptive methods. More recently, \cite{talili2013} considered a joint model with mixed correlated toxicity and  efficacy outcomes, with one discrete and the other continuous.  Using outcomes constructed with Archimedean Copula, they extended the continual reassessment method to find the optimal dose for Phase III trials based on both efficacy and toxicity considerations. Some advantages and disadvantages of adaptive designs have been discussed in \cite{detborbre2013}.

Recently, the   use of active controls instead of placebo in clinical trials has received considerable attention  in the literature [see
for example  \cite{pct}, \cite{ethics}, \cite{helbenfri2014,helbenzinknefri2015} among many others], and design issues for univariate
outcomes have been discussed in \cite{detkisbenbre2014,detketbre2015}. To our best knowledge,
the problem of determining optimal designs  for active controlled trials with bivariate mean outcomes has not been considered in the literature so far.
In the present paper we  derive optimal designs when the bivariate outcomes are efficacy and toxicity measures and are modeled using
  combinations of the linear, quadratic,  Michaelis-Menten and Emax model as  possible mean functions. 
 We note that the Emax model is especially flexible in its shape and commonly applied in dose-finding trials. In particular, the Emax model can be justified through the relationship of drug-receptor interactions
and therefore deduced from the chemical equilibrium equation [see e.g. \cite{borou2002}].

In Section \ref{sec2} we introduce the model and provide some
technical background.  In Section \ref{sec3}  we consider various models for efficacy and toxicity outcomes without an active control and provide
upper bounds on the required number of doses for each combination of the possible mean functions in the bivariate model.
We also  state sufficient conditions under which the boundary points of the design space are included 
as support points of  the optimal design and
determine the  minimally supported  optimal designs  analytically.
 In Section \ref{sec4} we apply these  results to design active-controlled dose finding trials. In Section \ref{sec5} we report the optimal designs for a real trial with bivariate outcome using particle swarm optimization. We conclude with a discussion in Section \ref{sec6} and provide the technical proofs for our main results in Section \ref{sec7}.

\section{Optimal designs for bivariate outcome}  \label{sec2}
\def\theequation{2.\arabic{equation}}
\setcounter{equation}{0}

We consider a dose-finding trial investigating both the efficacy and toxicity of a new drug under investigation. Our goal is to find an optimal design for collecting data for the two outcomes at different dose levels.  Given a statistical model defined on a  given dose interval of interest, say $\mathcal{D}= [L,R] \subset \R_0^+$,  and a given design criterion, the design problem is to determine the optimal number  of doses, $k$,  the dose levels $d_1, \ldots, d_k$ from the dose interval $\D$, and the number of patients assigned to  each dose. In practice, the total sample size, say $n_1$, is determined either by standard power considerations or by requirements on the precision of estimating the dose-response curves. That is, for a given value of $n_1$, the optimal design needs to specify the number of patients $n_{1i}$ at each dose level $d_i$ subject to $\sum_{i=1}^k n_{1i} = n_1$. Note that we use the index ``$1$'' here in the notation (i.e. $n_1,n_{1i},\ldots) $ since in later sections we consider active controlled clinical trials with a second sample denoted by the index ``$2$''.

Let  $Y_{ij} $ be the two-dimensional outcome variable at dose level $d_i$ from subject $j$ and assume that 
\begin{equation} \label{eq1}
Y_{ij}=(Y^e_{ij},Y^t_{ij})^T \sim \mathcal{N}_2(\eta_1(d_i,\theta_1),\Sigma_1), \qquad j=1,\dots,n_{1i}, i=1, \ldots, k.
\end{equation}
 The regression function
  $$
  \eta_1(d,\theta_1)=(\eta_1^e(d,\theta_1^e),\eta_1^t(d,\theta_1^t))^T \in \R^2
  $$
describes the expected efficacy ($\eta_1^e$) and toxicity ($\eta_1^t$) at dose level $d \in \D$, where the $(s_1^e+1)$- and $(s_1^t+1)$-dimensional vectors $\theta_1^e$ and $\theta_1^t$ define the parameters of the model $\eta_1^e$ and $\eta_1^t$, respectively.
The parameter $\theta_1=((\theta_1^e)^T,(\theta_1^t)^T)^T$  varies in a compact parameter space, say $\Theta_1 \subset \R^{s_1}$, where
$s_1=s_1^e+s_1^t+2$.  The unknown covariance matrix  is
\begin{equation*}
	\Sigma_1 = \Cov(Y) = \begin{pmatrix} \sigma_e^2  & \rho \sigma_e \sigma_t \\ \rho \sigma_e \sigma_t & \sigma_t^2  \end{pmatrix},
\end{equation*}
where $-1<\rho<1$ denotes  the correlation  between the two outcome variables and the variances  of the random variables $Y^e_{ij}$ and $Y^t_{ij}$
are given by $\sigma_e^2$ and $\sigma_t^2$, respectively.
We assume that $\eta_1$ is continuously differentiable with respect to the parameter $\theta_1$ and denote by
\begin{eqnarray*}
	f_{e}(d) &=& \frac{\partial}{\partial \theta_1^e} \eta_1^e(d,\theta_1)=(f_0^e(d),\ldots,f_{s_1^e}^e(d))^T ~,~
	f_{t}(d) = \frac{\partial}{\partial \theta_1^t} \eta_1^t(d,\theta_1)=(f_0^t(d),\ldots,f_{s_1^t}^t(d))^T
\end{eqnarray*}
the gradients of the two mean responses  with respect to  $\theta_1^e$ and $ \theta_1^t$, respectively.
Following \cite{fatu2001}, the Fisher information matrix is given by the $s_1 \times s_1$-matrix
\begin{eqnarray*}
	\I_1(d,\theta_1) &=& (\tfrac{\partial}{\partial \theta_1} \eta_1(d,\theta_1) )^T \Sigma_1^{-1} ( \tfrac{\partial}{\partial \theta_1} \eta_1(d,\theta_1))
	=\begin{pmatrix} f_{e}(d) & \mathbf{0}_{s_1^e+1  } \\ \mathbf{0}_{s_1^t+1  } & f_{t}(d) \end{pmatrix} \Sigma_1^{-1} \begin{pmatrix} f_{e}^T(d) & \mathbf{0}_{s_1^t+1}^T \\ \mathbf{0}_{ s_1^e+1}^T & f_{t}^T(d) \end{pmatrix}  \\
	&=& \frac{1}{\sigma_e^2\sigma_t^2(1-\rho^2)} F(d).
	\end{eqnarray*}
	Here, $\mathbf{0}_d$ is  the $d$-dimensional vector with all entries equal to $0$ and 
	\begin{eqnarray} \label{fmat}
F(d)&=&	\frac{1}{\sigma_e^2\sigma_t^2(1-\rho^2)} \begin{pmatrix} \sigma_t^2 \mathcal{F}_1 & -\rho\sigma_e \sigma_t \mathcal{F}_2 \\
-\rho\sigma_e \sigma_t \mathcal{F}_2^T & \sigma_e^2 \mathcal{F}_3 \end{pmatrix}
\end{eqnarray}
is defined through
\begin{eqnarray*}
	\mathcal{F}_1 &=& f_e(d) f_e^T(d) \in \R^{s_1^e+1 \times s_1^e+1} ,  \quad \mathcal{F}_3 = f_t(d) f_t^T(d) \in \R^{s_1^t+1 \times s_1^t+1}  , \\
	\mathcal{F}_2 &=& f_e(d) f_t^T(d)  \in \R^{s_1^e+1 \times s_1^t+1} .
\end{eqnarray*}
Note that we have suppressed the dependency of the matrices $F$, $\mathcal{F}_1$, $\mathcal{F}_2$ and $\mathcal{F}_3$ on the parameter $\theta_1$ in our notation.

Throughout this paper we consider approximate designs in the sense of \cite{kiefer1974}, which are defined as probability measures with finite support on the design space $\D$. If an approximate design $\xi$ has $k$ support points, say $d_1,\dots,d_k$,
 with corresponding positive weights $\omega_1, \ldots, \omega_k$,   such that $\sum_{i=1}^k \omega_i = 1$, and $n_1$ observations can be taken, a rounding procedure is applied to obtain integers $n_{1i}$, $i=1,\ldots,k$, from the possibly rational numbers $\omega_i n_1$ [see \cite{pukrie1992}].
The information matrix $M_1(\xi,\theta_1)$ of a design $\xi$ is defined by the $s_1 \times s_1$ matrix
\begin{equation}\label{Mone}
	M_1(\xi,\theta_1) =  \int_\D \I_1(d,\theta_1) d\xi(d) =  \sum_{i=1}^k  \frac{\omega_i}{\sigma_e^2\sigma_t^2(1-\rho^2)} F(d_i),
\end{equation}
%(\tfrac{\partial}{\partial \theta_1}\eta_1(d_i,\theta_1))^T\Sigma_1^{-1}(\tfrac{\partial}{\partial \theta_1}\eta_1(d_i,\theta_1))
where   the   matrix $F(d)$ is defined in \eqref{fmat}.

%\begin{footnotesize}
%\begin{equation*}
%\begin{pmatrix}
%		\sigma_t^2 (f_{0}^e(d))^2 & \ldots & \ldots & \ldots & \ldots  &\ldots&\ldots \\
%			\sigma_t^2 f_{0}^e(d)f_{1}^e(d) & \sigma_t^2 (f_{1}^e(d))^2 & \ldots & \ldots & \ldots & \ldots & \ldots \\
%		 \vdots & \vdots & \ddots & \ldots& \ldots & \ldots & \ldots  \\
%				\sigma_t^2 f_{0}^e(d)f_{{s_1^e}}^e(d) &  \sigma_t^2 f_{1}^e(d)f_{{s_1^e}}^e(d) & \hdots &  \sigma_t^2 (f_{{s_1^e}}^e(d))^2 & \ldots  & \ldots& \ldots \\
%				-\rho \sigma_e \sigma_t  f_{0}^e(d)f_{{0}}^t(d) & -\rho \sigma_e \sigma_t  f_{1}^e(d)f_{0}^t(d) & \hdots &  -\rho \sigma_e \sigma_t  f_{{s_1^e}}^e(d)f_{0}^t(d) & \sigma_e^2 (f_{0}^t(d))^2 & \ldots  & \ldots  \\
%							\vdots & \vdots  & \ldots  & \ldots   & \ldots  & \ddots & \ldots  &   \\
%								-\rho \sigma_e \sigma_t  f_{0}^e(d)f_{{s_1^t}}^t(d) & -\rho \sigma_e \sigma_t  f_{1}^e(d)f_{{s_1^t}}^t(d)& \ldots   & \hdots & \ldots    & \ldots  &  \sigma_e^2 (f_{{s_1^t}}^t(d))^2
%		\end{pmatrix}.
%\end{equation*}
%\end{footnotesize}

If observations are  taken according to an approximate design $\xi$
it can be shown that, under standard regularity conditions, the maximum likelihood estimator $\hat \theta_1$ is asymptotically normally distributed, that is
\begin{equation*}
	\sqrt{n_1}(\hat \theta_1 - \theta_1) \stackrel{ \mathcal{D}}{\longrightarrow} \mathcal{N}_{s_1}  (\mathbf{0}, M_1^{-1}(\xi,\theta_1)),
\end{equation*}
as $n_1 \to \infty$, where the symbol $\stackrel{ \mathcal{D}}{\longrightarrow}$ means convergence in distribution. Consequently 
designs  that make the information matrix $M_1(\xi,\theta_1)$ large in some sense are appropriate.
There are several design criteria used in practice.  An important example is Kiefer's $\phi_p$-criterion  [see \cite{kiefer1974}].  To be precise, let $ p \in [-\infty,1)$ and let  $ K \in \mathbb{R}^{s_1 \times m}$ be  a given matrix of full column rank. A design $\xi^*$  is called locally $\phi_p$-optimal for estimating the linear combination $K^T \theta_1$  if it maximizes the concave functional
\begin{equation*} %\label{critt}
\phi_p(\xi) =  \Bigl(\frac {1}{m} \tr(K^T M_1^- (\xi,\theta_1)K)^{-p} \Bigr)^{\frac {1}{p}}
\end{equation*}
among all designs $\xi$ satisfying Range$(K) \subset$ Range$(M_1(\xi,\theta_1))$, i.e. $K^T \theta_1$ is estimable by the design $\xi$. Here, $\tr(A)$ and $A^-$ denote the trace and a generalized inverse of the matrix $A$, respectively.

One key advantage of working with approximate designs is that convex optimization theory can be applied if the design criterion is a concave functional. As a consequence, a general equivalence theorem is available to verify whether a design is optimal  among all designs.  Any concave functional has its own
equivalence theorem but collectively they all have a similar form.  For each member of Kiefer's  $\phi_p$-criterion, a direct application of Theorem 7.14 in \cite{pukelsheim2006}   yields the following result.

 \begin{Theorem}\label{equithm}
  Let K be a $s_1\times m$ matrix of full column rank. If   $p \in (-\infty , 1)$, a design $\xi^*$ with  ${\rm} Range(K) \subset  {\rm}Range (M_1(\xi^*,\theta_1))$ is locally $\phi_p$-optimal for estimating the linear combination $K^T \theta_1$ if and only if there exists a generalized inverse $G$ of the   information matrix $M_1(\xi^*,\theta_1)$, such that
\begin{equation}\label{aequ}
\tr \bigl( \I_1(d,\theta_1)GK( C_K(\xi^*))^{p+1} K^TG^T\bigr) ~-~ \tr ( C_K(\xi^*))^{p}\leq 0
\end{equation}
holds for all $d \in \D$, where $C_K(\xi^*) =(K^TM^-_1(\xi^*,\theta_1)K)^{-1}$.
If $p = -\infty  $, a design $\xi^*$ with Range$(K) \subset$ Range$(M_1(\xi^*,\theta_1))$  is locally $\phi_{-\infty}$-optimal for estimating the linear combination $K^T \theta_1$ if and only if there exists a generalized inverse $G$ of the information matrix $M_1(\xi^*,\theta_1)$ and a non-negative definite matrix $E \in \mathbb{R}^{m \times m}$ with ${\rm tr} (E) =1$,
such that
\begin{equation}\label{aequ1}
 \tr \big( \I_1 (d,\theta_1) GK C_K(\xi^*)  E  C_K(\xi^*) K^TG^T\big)  - \lambda_{\min}(C_K(\xi^*)) \leq 0.
\end{equation}
holds for all $d \in \mathcal{D}$. Moreover, if the design $\xi^*$ is $\phi_p$-optimal, there is equality in the above inequalities.
 \end{Theorem}

The function on the left hand side of (\ref{aequ}) or (\ref{aequ1}) is a function of the dose $d$ and is sometimes called the sensitivity function of the design $\xi^*$.  In practice, one plots the sensitivity function over the entire dose range and checks whether it is bounded above by zero.
%with equality at the support points of $\xi^*$.
  If it does, the design $\xi^*$ is optimal; otherwise it is not.  In addition, the sensitivity function, along with the equivalence theorem, can be used to provide a lower bound on the efficiency of any design. For example, if $p > - \infty$, one can show that the $\phi_p$-efficiency
  of a design $\xi$  can be bounded from below, that is
  \begin{equation} \label{effbound}
  \mbox{eff}_p(\xi) = \frac {\phi_p(\xi)}{\sup_\eta \phi_p(\eta)} \geq \frac { \tr ( C_K(\xi))^{p}}  {\max_{d \in  \mathcal{D}} \tr  ( \I_1(d,\theta_1)GK( C_K(\xi))^{p+1} K^TG^T) }
  \end{equation}
  [see \cite{dett:1996}].
  Moreover, characterizations of the type \eqref{aequ} or \eqref{aequ1} are also useful to
   find optimal designs analytically
   %and provide good information on the number and distribution of  support points
   if the model is not too complicated. However, regression models with multivariate outcome
   are  complex and in practice optimal designs have to be found numerically  [see \cite{chang1997}, \cite{atasei2007} or \cite{sagnol2011} among others]. \\
   \noindent
 For such calculations, sharp bounds on the number of support points of the optimal designs are useful, because they can substantially reduce the complexity of the optimization problem.

In order to derive such upper bounds we follow  \cite{karstud1966} and call a design $\xi_1$ admissible if there does not exist a design $\xi_2$, such that $M_1(\xi_1,\theta_1) \neq M_1(\xi_2,\theta_1)$ and
\begin{equation*} %\label{admissible}
	M_1(\xi_1,\theta_1) \leq M_1(\xi_2,\theta_1)
\end{equation*}
with respect to the Loewner ordering. In other words, the information matrix of an admissible design cannot be improved and numerical optimization can be restricted to the class of admissible designs.
The characterization of the number of support points of admissible designs has found considerable attention in the recent literature [see \cite{yangstuf2009,yangstuf2012b}, \cite{yang2010}, \cite{detmel2011} or \cite{dettscho2013}]. These authors obtained  substantially smaller bounds on the number of support points of optimal
designs than provided by the classical approach using Caratheodory's theorem [see \cite{pukelsheim2006} for example].

In the following we  demonstrate that the results in the  above references can in fact be proved under weaker assumptions than usually made
in the literature. For this purpose
we will make use of the theory of Tchebycheff systems [see \cite{karstud1966}].
A  set of $k+1$ continuous functions $u_0,\ldots,u_k \colon [L,R] \to \R$ is called a Tchebycheff system
on the interval $[L,R]$  if the inequality
$\det (u_i(d_j))^k_{i,j=0}>0$ holds for all $L \leq d_0 < d_1 < \ldots < d_k \leq R$. We  define the index $I(\xi)$ of a design $\xi$ on the interval $[L,R]$  as the number of support points, where interior
support points are counted by one and the support points at the boundary  of the interval $[L,R]$ are counted by one half.

Note that we can rewrite the information matrix $M_1(\xi,\theta_1)$ in the form
\begin{equation} \label{Monepsi}
	M_1(\xi,\theta_1) = \begin{pmatrix} \int_L^R \psi_{11}(d)d\xi(d) & \ldots & \int_L^R \psi_{1s_1}(d)d\xi(d) \\ \vdots & & \vdots \\ \int_L^R \psi_{s_11}(d)d\xi(d) & \ldots & \int_L^R \psi_{s_1s_1}(d)d\xi(d) \end{pmatrix},
\end{equation}
where we ignore the dependence of the functions $\psi_{ij}$ on the parameter $\theta_1$.
We now define $\psi_0(d) \equiv 1$ and choose a basis, say  $\{ \psi_0,\ldots,\psi_k \}$, for the space
 ${\rm span} (\{\psi_{ij} | 1 \leq i,j \leq s_1  \} \cup \{ 1\}  )$. We further
assume   that $\psi_k$ is one of the diagonal elements of the matrix $M_1(\xi,\theta_1)$, does not coincide with any of the other elements $\psi_{ij}$
 and that $\{ \psi_0,\ldots,\psi_{k-1} \}$ is a basis of the space
$$
\mbox{span}\big(  \{ \psi_{ij} \mid i,j \in \{1,\ldots,s_1\}; \ \ \psi_{ij} \neq \psi_k \} \cup \{1\}  \big) .
$$
Our next result, Theorem \ref{detmelmod},  is  a more general version of Theorem $3.1$ in \cite{detmel2011} that is specific to our problem here. The proof
is quite similar to the one given in this reference and is omitted for the sake of brevity. Theorem \ref{detmelmod} yields 
better  bounds on the number of support points of  optimal designs obtained from the current literature; an example 
is given at the end of Section
\ref{sec71}.

\newpage
\begin{Theorem} \label{detmelmod} ~
	\begin{itemize}
		\item[(A)] If $\{\psi_0,\psi_1,\ldots,\psi_{k-1}\} $ {and} $ \{\psi_0, \psi_1,\ldots,\psi_{k}\}$ are  Tchebycheff systems on the interval $\D$,
		 then for any design $\xi$ there exists a design $\xi^+$ with at most $\tfrac{k+2}{2}$ support points, such that $M_1(\xi^+,\theta_1) \geq M_1(\xi,\theta_1)$. If the index of the design $\xi$ satisfies $I(\xi) < \frac{k}{2}$,
		 then the design $\xi^+$ is uniquely determined in the class of all designs $\eta$ satisfying
		\begin{equation} \label{bed}
			\int_L^R \psi_i(d) d\eta(d) = \int_L^R \psi_i(d)d\xi(d), \quad i=0,\ldots,k-1
		\end{equation}
		and coincides with the design $\xi$. Otherwise, in the case $I(\xi)\geq\tfrac{k}{2}$, the following two assertions are valid.
		\begin{enumerate}
			\item[(A1)] If $k$ is odd, then $\xi^+$ has at most $\tfrac{k+1}{2}$ support points and $\xi^+$ can be chosen such that its support contains the point $R$.
			\item[(A2)] If $k$ is even, then $\xi^+$ has at most $\tfrac{k}{2}+1$ support points and $\xi^+$ can be chosen such that its support contains the points $L$ and $R$.
		\end{enumerate}
		
		\item[(B)] If $\{\psi_0,\psi_1,\ldots,\psi_{k-1}\} $ {and} $ \{\psi_0, \psi_1,\ldots,-\psi_{k}\} $ are  Tchebycheff systems,
		then for any design $\xi$ there exists a design $\xi^-$ with at most $\tfrac{k+2}{2}$ support points, such that $M_1(\xi^-,\theta_1) \geq M_1(\xi,\theta_1)$. If the index of the design $\xi$ satisfies $I(\xi) < \frac{k}{2}$
then the design $\xi^-$ is uniquely determined in the class of all designs $\eta$ satisfying \eqref{bed} and coincides with the design $\xi$. Otherwise, in the case $I(\xi)\geq\tfrac{k}{2}$, the following two assertions are valid.
		\begin{enumerate}
			\item[(B1)] If $k$ is odd, then $\xi^-$ has at most $\tfrac{k+1}{2}$ support points and $\xi^-$ can be chosen such that its support contains the point $L$.
			\item[(B2)] If $k$ is even, then $\xi^-$ has at most $\tfrac{k}{2}+1$ support points.
		\end{enumerate}
	\end{itemize}
\end{Theorem}

We note that    Theorem \ref{detmelmod} provides information about the admissible designs. For example, consider the case  $(A2)$ with $k=2m$ for some $m \in \mathbb{N}$. Any design $\xi$ with index $I(\xi) \geq m$ can be improved with respect to the Loewner ordering by a design  with at most $m+1$ support 
points that includes the boundary points $L$ and $R$. It follows that  admissible designs are  designs with index $<m$ and designs with $m+1$ support points that include the boundary points $L$ and $R$ of the design space.

%  In Section \ref{prel} we will formulate a slightly more general version of Theorem $3.1$ in \cite{detmel2011}
%  specific to our problem, which characterizes the admissible designs under certain assumptions on the entries of
%  the information matrix.

%%%%%%%%%------------------------------------%%%%%%%%%%%%%%%%%%%%%%%%%%%%%%%%

\section{Optimal designs for placebo-controlled dose finding trials} \label{sec3}
\def\theequation{3.\arabic{equation}}
\setcounter{equation}{0}

In this section we study optimal designs for several nonlinear regression models which are commonly used in 
placebo-controlled dose-finding trials with joint efficacy-toxicity outcomes.
In particular we  use Theorem \ref{detmelmod} to derive bounds on the number of support points of optimal designs and explicit
expressions for minimally supported designs. The proofs of the results presented here can be found in the Appendix.

\subsection{Bounds on the number of support points}
In order to determine bounds for the number of support points of optimal designs we note that the mapping $M \to (K^TM^-K)^{-1}$ is increasing with respect to the Loewner ordering on the set of all $s_1 \times s_1$-matrices satisfying Range$(K) \subset$ Range$(M)$ [see \cite{pukelsheim2006}]. That is, if
\begin{equation*}
	M_1 \geq M_2   \quad \Rightarrow \quad (K^TM_1^-K)^{-1} \geq (K^TM_2^-K)^{-1},
	\end{equation*}
for all matrices $M_1, M_2$ satisfying the range inclusion.
It therefore follows that the information matrix $(K^TM^-(\xi,\theta_1)K)^{-1}$ of a non-admissible design can be improved  with respect to the Loewner ordering.
Because the $\phi_p$-criteria are monotone, we have $\phi_p(\xi) \leq \phi_p(\xi^*)$
for any design $\xi$, where $\xi^*$ is either $\xi^+$ or $\xi^-$ as given in Theorem \ref{detmelmod}.
This conclusion is true, whenever the assumptions of Theorem \ref{detmelmod} are satisfied.  The following results show that this is in fact the case for many of the commonly used dose response models with a bivariate outcome and give  upper bounds on the number of support points of such designs.

\begin{Theorem} \label{thmefftoxLIN}
Assume that the model for efficacy is given by $\eta_1^e(d,\theta_1)=\vartheta_0^e + \vartheta_1^e d$ and that $\xi$ 
is  an arbitrary design on the dose range ${\cal D} = [L,R]$.
\begin{itemize}
\item[(a)]
If $\eta_1^t(d,\theta_1)=\vartheta_0^t + \vartheta_1^t d$,  there exists a design $\xi^*$ with at most two support points, such that $M_1(\xi^*,\theta_1)\geq M_1(\xi,\theta_1)$. If the index of the design $\xi$ satisfies $I(\xi)\geq 1$,  $\xi^*$ can be chosen such that the support of $\xi^*$ contains the points $L$ and $R$.
%\item[(b)]
%If $\eta_1^t(d,\theta_1)=\vartheta_0^t \exp(\tfrac{d}{\vartheta_1^t})$, then there exists a design $\xi^*$ with at most five support points, such that $M_1(\xi^*,\theta_1)\geq M_1(\xi,\theta_1)$. If the index of the design $\xi$ satisfies $I(\xi)\geq 4$, then $\xi^*$ can be chosen such that the support of $\xi^*$ contains $L$ and $R$.
\item[(b)] If $\eta_1^t(d,\theta_1)=\vartheta_0^t + \vartheta_1^t d + \vartheta_2^t d^2$, 
 there exists a design $\xi^*$ with at most three support points, such that $M_1(\xi^*,\theta_1)\geq M_1(\xi,\theta_1)$. If the index of the design $\xi$ satisfies $I(\xi)\geq 2$,  $\xi^*$ can be chosen such that the support of $\xi^*$ contains the points $L$ and $R$.
\item[(c)] If $\eta_1^t(d,\theta_1)$ is given by a Michaelis-Menten model, that is $\eta_1^t(d,\theta_1)=\tfrac{\vartheta_1^t d}{\vartheta_2^t + d}$,  there exists a design $\xi^*$ with at most four support points, such that $M_1(\xi^*,\theta_1)\geq M_1(\xi,\theta_1)$. If the index of the design $\xi$ satisfies $I(\xi)\geq 3$, 
 $\xi^*$ can be chosen such that the support of $\xi^*$ contains the points $L$ and $R$.
\item[(d)] If $\eta_1^t(d,\theta_1)$ is given by an Emax-model, that is $\eta_1^t(d,\theta_1)=\vartheta_0^t +\tfrac{\vartheta_1^t d}{\vartheta_2^t + d}$, 
 there exists a design $\xi^*$ with at most four support points, such that $M_1(\xi^*,\theta_1)\geq M_1(\xi,\theta_1)$. If the index of the design $\xi$ satisfies $I(\xi)\geq 3$,  $\xi^*$ can be chosen such that the support of $\xi^*$ contains the points $L$ and $R$.
\end{itemize}
\end{Theorem}

\begin{Remark} \label{rem0}
{\rm Note that the bounds provided by Theorem \ref{thmefftoxLIN} are not necessarily sharp. For example, if $\eta^t_1$
is  the Emax and $\eta^e_1$  is the linear model, then by the first part of Theorem \ref{thmefftoxLIN}(d) one does not decrease the information (with respect to the L{oe}wner ordering) by considering only designs with at most four support points. Any design with four support points or three support points in the interior of the dose range has index $\geq 3$ and can therefore be further improved by a design with at most four support points including the boundary points $L$ and $R$. As one requires at least three different dose levels to estimate the parameters  in the Emax model, it follows that one can restrict the search of optimal designs to three point designs with at least one boundary point as support point (as the index should be smaller than or equal to $5/2$) or to four point designs containing both boundary points in its support.}
\end{Remark}

\begin{Theorem} \label{thmefftoxQUAD}
Assume that the model for efficacy is given by $\eta_1^e(d,\theta_1)=\vartheta_0^e + \vartheta_1^e d + \vartheta_2^e d^2$ and let $\xi$ denote
an arbitrary design on the dose range ${\cal D} = [L,R]$.
\begin{itemize}
%\item[(a)] If $\eta_1^t(d,\theta_1)=\vartheta_0^t + \vartheta_1^t d$, then there exists a design $\xi^*$ with at most three support points, such that $M_1(\xi^*,\theta_1)\geq M_1(\xi,\theta_1)$. If the index of the design $\xi$ satisfies $I(\xi)\geq 2$, then $\xi^*$ can be chosen such that the support of $\xi^*$ contains the points $L$ and $R$.
\item[(a)] If $\eta_1^t(d,\theta_1)=\vartheta_0^t + \vartheta_1^t d+\vartheta_2^t d^2$,  there exists a design $\xi^*$ with at most three support points, such that $M_1(\xi^*,\theta_1)\geq M_1(\xi,\theta_1)$. If the index of the design $\xi$ satisfies $I(\xi)\geq 2$, 
 $\xi^*$ can be chosen such that the support of $\xi^*$ contains the points $L$ and $R$.
\item[(b)] If $\eta_1^t(d,\theta_1)=\tfrac{\vartheta_1^t d}{\vartheta_2^t + d}$,  there exists a design $\xi^*$ with at most five support points, such that $M_1(\xi^*,\theta_1)\geq M_1(\xi,\theta_1)$. If the index of the design $\xi$ satisfies $I(\xi)\geq 4$, 
 $\xi^*$ can be chosen such that the support of $\xi^*$ contains the points $L$ and $R$.
\item[(c)] If $\eta_1^t(d,\theta_1)=\vartheta_0^t + \tfrac{\vartheta_1^t d}{\vartheta_2^t + d}$,  there exists a design $\xi^*$ with at most five support points, such that $M_1(\xi^*,\theta_1)\geq M_1(\xi,\theta_1)$. If the index of the design $\xi$ satisfies $I(\xi)\geq 4$,  $\xi^*$ can be chosen such that the support of $\xi^*$ contains  the points $L$ and $R$.
\end{itemize}
\end{Theorem}

\begin{Theorem} \label{thmefftoxMM}
Assume that the model for efficacy is given by $\eta_1^e(d,\theta_1)=\tfrac{\vartheta_1^e d}{\vartheta_2^e + d}$ and let $\xi$ denote
an arbitrary design on the dose range ${\cal D} = [L,R]$.
\begin{itemize}
%\item[(a)]
%If $\eta_1^t(d,\theta_1)=\vartheta_0^t + \vartheta_1^t d$, then there exists a design $\xi^*$ with at most four support points, such that $M_1(\xi^*,\theta_1)\geq M_1(\xi,\theta_1)$. If the index of the design $\xi$ satisfies $I(\xi)\geq 3$, then $\xi^*$ can be chosen such that the support of $\xi^*$ contains the points $L$ and $R$.
%\item[(b)] If $\eta_1^t(d,\theta_1)=\vartheta_0^t + \vartheta_1^t d+\vartheta_2 d^2$, then there exists a design $\xi^*$ with at most five support points, such that $M_1(\xi^*,\theta_1)\geq M_1(\xi,\theta_1)$. If the index of the design $\xi$ satisfies $I(\xi)\geq 4$, then $\xi^*$ can be chosen such that the support of $\xi^*$ contains the points $L$ and $R$.
%\item[(b)] EXP \\
%If $\eta_1^t(d,\theta_1)=\vartheta_0^t \exp(\tfrac{d}{\vartheta_1^t})$, then there exists a design $\xi^*$ with at most \textcolor{red}{X} support points, such that $M_1(\xi^*,\theta_1)\geq M_1(\xi,\theta_1)$. If the index of the design $\xi$ satisfies $I(\xi)\geq \textcolor{red}{X}$, then $\xi^*$ can be chosen such that the support of $\xi^*$ contains \textcolor{red}{X}.
\item[(a)] If $\eta_1^t(d,\theta_1)=\tfrac{\vartheta_1^t d}{\vartheta_2^t + d}$ with $\vartheta_2^e \neq \vartheta_2^t$,  there exists a design $\xi^*$ with at most five support points, such that $M_1(\xi^*,\theta_1)\geq M_1(\xi,\theta_1)$. If the index of the design $\xi$ satisfies $I(\xi)\geq 4$, 
 $\xi^*$ can be chosen such that the support of $\xi^*$ contains the point  $R$.
\item[(b)] If $\eta_1^t(d,\theta_1)=\vartheta_0^t+\tfrac{\vartheta_1^t d}{\vartheta_2^t + d}$ with $\vartheta_2^e \neq \vartheta_2^t$,  there exists a design $\xi^*$ with at most five support points, such that $M_1(\xi^*,\theta_1)\geq M_1(\xi,\theta_1)$. If the index of the design $\xi$ satisfies $I(\xi)\geq 4$,  $\xi^*$ can be chosen such that the support of $\xi^*$ contains the points $L$ and $R$.
\end{itemize}
\end{Theorem}

\begin{Theorem} \label{thmefftoxEMAX}
Assume that the model for efficacy is given by $\eta_1^e(d,\theta_1)=\vartheta_0^e + \tfrac{\vartheta_1^e d}{\vartheta_2^e + d}$ and let $\xi$ denote
an arbitrary design on the dose range ${\cal D} = [L,R]$.
%\item[(a)] If $\eta_1^t(d,\theta_1)=\vartheta_0^t + \vartheta_1^t d$, then there exists a design $\xi^*$ with at most four support points, such that $M_1(\xi^*,\theta_1)\geq M_1(\xi,\theta_1)$. If the index of the design $\xi$ satisfies $I(\xi)\geq 3$, then $\xi^*$ can be chosen such that the support of $\xi^*$ contains the points $L$ and $R$.
%\item[(b)] If $\eta_1^t(d,\theta_1)=\vartheta_0^t \exp(\tfrac{d}{\vartheta_1^t})$, then there exists a design $\xi^*$ with at most five support points, such that $M_1(\xi^*,\theta_1)\geq M_1(\xi,\theta_1)$. If the index of the design $\xi$ satisfies $I(\xi)\geq 4$, then $\xi^*$ can be chosen such that the support of $\xi^*$ contains the points $L$ and $R$.
%\item[(b)] If $\eta_1^t(d,\theta_1)=\vartheta_0^t + \vartheta_1^t d+\vartheta_2^t d^2$, then there exists a design $\xi^*$ with at most five support points, such that $M_1(\xi^*,\theta_1)\geq M_1(\xi,\theta_1)$. If the index of the design $\xi$ satisfies $I(\xi)\geq 4$, then $\xi^*$ can be chosen such that the support of $\xi^*$ contains the points $L$ and $R$.
%\item[(c)] If $\eta_1^t(d,\theta_1)=\tfrac{\vartheta_1^t d}{\vartheta_2^t + d}$, then there exists a design $\xi^*$ with at most five support points, such that $M_1(\xi^*,\theta_1)\geq M_1(\xi,\theta_1)$. If the index of the design $\xi$ satisfies $I(\xi)\geq 4$, then $\xi^*$ can be chosen such that the support of $\xi^*$ contains the points $L$ and $R$.
If $\eta_1^t(d,\theta_1)=\vartheta_0^t + \tfrac{\vartheta_1^e d}{\vartheta_2^e + d}$ with $\vartheta_2^e \neq \vartheta_2^t$,  there exists a design $\xi^*$ with at most five support points, such that $M_1(\xi^*,\theta_1)\geq M_1(\xi,\theta_1)$. If the index of the design $\xi$ satisfies $I(\xi)\geq 4$,  $\xi^*$ can be chosen such that the support of $\xi^*$ contains the points $L$ and $R$.
\end{Theorem}

\medskip

\begin{Remark} \label{rem1}
{\rm  The remaining cases can be obtained by interchanging the roles of $\eta^e$ and $ \eta^t$
in Theorem  \ref{thmefftoxLIN} -  \ref{thmefftoxEMAX}. For example, consider the case, where
$\eta_1^e(d,\theta_1)$ is the Emax and  $\eta_1^t(d,\theta_1)$
the Michaelis-Menten model with $\vartheta_2^e \neq \vartheta_2^t$.  Then it follows  from Theorem \ref{thmefftoxMM}(b) that for any design $\xi$
there exists a design $\xi^*$ with at most five support points, such that $M_1(\xi^*,\theta_1)\geq M_1(\xi,\theta_1)$.
Moreover, if the index of the design $\xi$ satisfies $I(\xi)\geq 4$, then $\xi^*$ can be chosen such that the support of $\xi^*$ contains $L$ and $R$.
The other cases are obtained in the same way.
}
\end{Remark}

\subsection{Minimally supported $D$-optimal designs} \label{sec32}
For a design $\xi$ let  $\# \, \rm{supp}(\xi)$ be  the number of its support points and let
$$
m^* = \min \{ \# \,{ \rm supp} (\eta) \mid \det (M_1(\eta,\theta_1)) > 0, \  \eta \   \mbox{design on} \  \mathcal{D}\}
$$
be  the minimal number of support points required for a design with a non-singular information matrix in model \eqref{eq1}.
A design $\xi$ is called minimally supported if $\det (M_1(\xi,\theta_1))>0$ and the number of support points is given by $m^*$. Minimally supported designs are useful if, for example,
a drug under investigation may  be   only available  at few dose levels. \\
 In general,  the optimal designs have to be found numerically for complex models and even then many of the current algorithms may not work well.  However, if one restricts the search to minimally supported designs, the optimization problem can be greatly simplified which may then allow us to determine locally
  $D$-optimal designs. In some cases these minimally supported optimal designs may not be optimal among all designs [see Section \ref{sec5} below for some examples] so that an equivalence theorem must be used to confirm its optimality among all designs or its efficiency should be evaluated using the estimate \eqref{effbound}. Before we present analytically derived minimally supported designs for model \eqref{eq1} for different efficacy-toxicity regression models, we give a result about the general structure of these designs.  
\begin{Theorem}\label{equallyweighted}
%Let the dimension of the parameter vectors $\theta_1^e$ and $\theta_1^t$  in model (1)   be equal, that is 
If  the number of parameters  in  the  mean function of  the efficacy model is the same as the number of parameters in 
 the mean function of the toxicity model, i.e. 
$s^e_1 = s^t_1$, the minimally supported locally D-optimal design for model  \eqref{eq1} is a uniform design. Moreover, its support points do not depend on the entries 
in  the covariance matrix $\Sigma_1$.
\end{Theorem}

The following result provides minimally supported $D$-optimal designs for several commonly used dose-response models. Its proof makes use of  Theorem \ref{equallyweighted}, which reduces the optimization problem to the determination of the support points.

\begin{Theorem} \label{thmefftoxminiLIN} ~~

\begin{itemize}
\item[\bf{(1)}] Assume that the model for efficacy is given by $\eta_1^e(d,\theta_1)=\vartheta_0^e + \vartheta_1^e d$.
\begin{itemize}
\item[(1a)] If $\eta_1^t(d,\theta_1)=\vartheta_0^t + \vartheta_1^t d$,  the minimally supported D-optimal design is a two-point design with equal masses at the  points $L$ and $R$.
\item[(1b)] If $\eta_1^t(d,\theta_1)=\tfrac{\vartheta_1^t d}{\vartheta_2^t + d}$,  the minimally supported D-optimal design is a two-point design with equal masses at the points $L \lor \frac{1}{2}(\sqrt{R^2+10 R \vartheta_2^t+9 (\vartheta_2^t)^2}-R-3 \vartheta_2^t)$ and $R$.
\end{itemize}
%\begin{Theorem} \label{thmefftoxminiQUAD}
\item[\bf{(2)}] Assume that the model for efficacy is given by $\eta_1^e(d,\theta_1)=\vartheta_0^t + \vartheta_1^t d+\vartheta_2^t d^2$.
\begin{itemize}
\item[(2a)] \textbf{}If $\eta_1^t(d,\theta_1)=\vartheta_0^t + \vartheta_1^t d+\vartheta_2^t d^2$,  the minimally supported D-optimal design is a three-point design with equal masses at the points $L$, $\tfrac{L+R}{2}$ and $R$.
\item[(2b)] If $\eta_1^t(d,\theta_1)=\vartheta_0^t + \tfrac{\vartheta_1^t d}{\vartheta_2^t + d}$,  the minimally supported D-optimal design is a three-point design with equal masses at the points $L$, $\sqrt{(L+\vartheta_2^t) (R+\vartheta_2^t)}-\vartheta_2^t$ and $R$.
\end{itemize}
% \begin{Theorem} \label{thmefftoxminiMM}
\item[\bf{(3)}]  Assume that the model for efficacy is given by $\eta_1^e(d,\theta_1)=\tfrac{\vartheta_1^e d}{\vartheta_2^e + d}$.
\begin{itemize}
\item[(3a)] If $\eta_1^t(d,\theta_1)=\vartheta_0^t + \vartheta_1^t d$,  the minimally supported D-optimal design is a two-point design with equal masses at the points $L \lor \frac{1}{2}(\sqrt{R^2+10 R \vartheta_2^e+9 (\vartheta_2^e)^2}-R-3 \vartheta_2^e)$ and $R$.
\item[(3b)]
If $\eta_1^t(d,\theta_1)=\tfrac{\vartheta_1^t d}{\vartheta_2^t + d}$,  the minimally supported D-optimal design is a two-point design with equal masses at the optimal points $L \lor \frac{\sqrt{ R \vartheta_2^e \vartheta_2^t (R+\vartheta_2^e+\vartheta_2^t)+ (\vartheta_2^e\vartheta_2^t)^2}- \vartheta_2^e \vartheta_2^t}{(R+\vartheta_2^e+\vartheta_2^t)}$ and $R$.
\end{itemize}
% \begin{Theorem} \label{thmefftoxminiEMAX}
\item[\bf{(4)}] Assume that the model for efficacy is given by $\eta_1^e(d,\theta_1)=\vartheta_0^e + \tfrac{\vartheta_1^e d}{\vartheta_2^e + d}$.
\begin{itemize}
\item[(4a)] If $\eta_1^t(d,\theta_1)=\vartheta_0^t + \vartheta_1^t d+\vartheta_2^t d^2$,  the minimally supported D-optimal design is a three-point design with equal masses at the points $L$, $\sqrt{(L+\vartheta_2^e) (R+\vartheta_2^e)}-\vartheta_2^e$ and $R$.
\item[(4b)]
If $\eta_1^t(d,\theta_1)=\vartheta_0^t + \tfrac{\vartheta_1^t d}{\vartheta_2^t + d}$,  the minimally supported D-optimal design is a three-point design with equal masses at the points $L$, $\frac{\sqrt{(L+\vartheta_2^e) (L+\vartheta_2^t) (R+\vartheta_2^e) (R+\vartheta_2^t)}+L R-\vartheta_2^e \vartheta_2^t}{L+R+\vartheta_2^e+\vartheta_2^t}$ and $R$.
\end{itemize}
\end{itemize}
\end{Theorem}

%%%%%%%%%------------------------------------%%%%%%%%%%%%%%%%%%%%%%%%%%%%%%%%
\section{Active-controlled dose-finding  trials} \label{sec4}
\def\theequation{4.\arabic{equation}}
\setcounter{equation}{0}

The use of active controls instead of placebo in clinical trials has received considerable attention  in the literature [see \cite{pct} and \cite{ethics} among many others].
In active controlled dose-finding trials patients are randomized to receive either one of several doses of the new drug or an active control 
(a marketed drug
administered at a specific dose level).
Inference issues for  active-controlled dose-finding trials were investigated only more recently [see, for example,
%Methods for analyzing such trials are available and they date back more than a decade old, see for example \cite{fleming,simon1999,compare},
%with more recent ones that include
\cite{helbenfri2014,helbenzinknefri2015}].
\cite{detkisbenbre2014,detketbre2015} investigated  optimal design problems  for such trials 
by determining  the optimal number of different dose levels, the individual dose levels within the dose range under investigation and the allocation ratios of patients at each dose level and the active control.
Despite the increasing importance of such trials [see \cite{compare}], there is virtually no work on developing optimal designs for active-controlled dose-finding trials with efficacy-toxicity outcomes, especially given the fact that designs for placebo controlled trials do not extend directly to active-controlled trials [see \cite{detkisbenbre2014}].
%Simialr A situation was similarly noted by \cite{sjwang} who expressed cautionary remarks that methods for analyzing placebo controlled trials cannot straightforwardly apply to active %controlled trials.

Our goal in this section is  to design an active-controlled dose-finding trial with a pre-determined total number of patients $N$ by determining the optimal number $k$ of different dose levels for the new drug, their individual dose levels $d_1,\ldots, d_k$, and  the optimal number $n_1$ of patients to be assigned to the new drug, along with the allocation scheme across the recommended doses.
The remaining number   $n_2 = N-n_1$ of patients are assigned to the the active control, which is assumed to be available at a fixed dose level $C$. In terms of approximate designs, we have designs of the form
\begin{equation} \label{xitilde}
 \tilde \xi = \begin{pmatrix} (d_1,0) & \ldots & (d_k,0) & (C,1) \\ \tilde\omega_1 & \ldots & \tilde\omega_k & \tilde\omega_{k+1} \end{pmatrix}~,
\end{equation}
where $\tilde\omega_i$ denotes the  proportion of
patients assigned treated at the $i^{th}$ dose level of the new drug, $i=1,\ldots,k$ and $ \tilde\omega_{k+1}$
the proportion of patients treated with the active control, that is $n_2 \approx \tilde \omega_{k+1}N$.
Here the second component of a design points in \eqref{xitilde} specifies if patients receive the new drug (``$0$'') or the active control
(``$1$'').
Note that the approximate design  $\tilde \xi $ induces an  approximate design of the form
\begin{equation} \label{xi}
	\xi = \begin{pmatrix} d_1 & \ldots & d_k \\ \omega_1 & \ldots & \omega_k \end{pmatrix},
\end{equation}
for the new drug  defining $\omega_i = \tilde \omega_i  / (1-\tilde\omega_{k+1})$.
Extending the statistical model from
\cite{detkisbenbre2014} to the efficacy-toxicity outcomes considered here, we have
\begin{eqnarray} \label{eq2}
Y_{ij} =(Y_{ij} ^e,Y_{ij} ^t)^T \sim \mathcal{N}_2(\eta_1(d_i,\theta_1),\Sigma_1) ~; j=1,\ldots , n_{1i} ,\\
Z_j =(Z_j^e,Z_j^t)^T \sim \mathcal{N}_2(\eta_2(\theta_2),\Sigma_2)~;j=1,\ldots ,n_2,
\label{eq2a}
\end{eqnarray}
where $Y_{ij}$ denotes the outcome from the $j$th patient  treated with the new drug at dose level $d_i$, and $Z_j$ the outcome from the $j$th patient  treated with the active control.
The  two-dimensional vector  $\eta_2 (\theta_2) $  is  the expected outcome,  and $\theta_2 $  a parameter which varies in a  compact  parameter space,
say  $\Theta_2$, and $\Sigma_2$ is a  $2 \times 2$ covariance matrix. The function $\eta_2: \Theta_2 \to \R^2$ is
 assumed to be continuously differentiable. Assuming that all observations are   independent, it can be  shown that the information matrix of a design $\tilde \xi$ defined in \eqref{xitilde} has a block-structure of the form
\begin{equation} \label{M}
	M(\tilde\xi, \theta) = \begin{pmatrix} (1-\tilde\omega_{k+1})M_1(\xi,\theta_1) & \bf{0} \\ \bf{0} & \tilde\omega_{k+1} \I_2(\theta_2)\end{pmatrix},
\end{equation}
where $\theta=(\theta_1^T,\theta_2^T)^T$ and
\begin{equation*}
		\I_2(\theta_2) = (\tfrac{\partial}{\partial \theta_2} \eta_2(\theta_2) )^T \Sigma_2^{-1} ( \tfrac{\partial}{\partial \theta_2} \eta_2(\theta_2))
\end{equation*}
is the Fisher information matrix corresponding to the active control.
Following \cite{detketbre2015} locally optimal designs for active-controlled dose-finding trials can be obtained from locally optimal designs for ordinary dose-finding trials. We extend this result to the class of admissible designs in Theorem \ref{thethmadm}, whose
 proof can be found in the Appendix.

\begin{Theorem} \label{thethmadm}
If $\xi$ is an admissible design of the form \eqref{xi} in model \eqref{eq1} and $\tilde\omega_{k+1} \in (0,1)$,  the design $\tilde \xi$ defined in \eqref{xitilde} is an admissible design for the model \eqref{eq2} with an active control \eqref{eq2a}.
\end{Theorem}

We now   characterize admissible designs for various regression functions in the model \eqref{eq2} with an active control. For this purpose we apply  Theorem \ref{thethmadm} to the results from Section \ref{sec3}. We illustrate the  methodology in an example with the Michaelis-Menten and Emax model. The other models discussed in Section \ref{sec3} can be considered in a similar way.

\begin{Example} \label{exefftoxAC}
{\rm Suppose that   the mean  outcome for toxicity is  given by an Emax model.  We consider two situations, where the efficacy outcome is first modeled by an Emax model and in the second case, is modeled by the Michaelis-Menten model.  For the first case, it  follows from Theorem \ref{thmefftoxEMAX} (d) that admissible designs in trials without an active control have at most five support points. By Theorem \ref{thethmadm}, we conclude that admissible designs in active-controlled trials are of the form \eqref{xitilde} with at most six support points and a positive weight $\tilde \omega_6 \in (0,1)$ for   the active control. Similarly, 
for the second case, it follows from Theorem \ref{thmefftoxMM} (b) that there exists an admissible design for the corresponding active-controlled trial with at most six support points with a positive weight $\tilde \omega_6 \in (0,1)$ for the active control. Moreover, the dose levels for the new drug include the boundary points $L$ and $R$ of the dose range.}
\end{Example}

In a similar way,  $\phi_p$-optimal designs for active-controlled trials with efficacy-toxicity outcomes can be obtained.
For this purpose we state the following result which can be proved in a similar way as Theorem $1$
 in \cite{detketbre2015} using the   block-structure of the matrix $M(\tilde \xi,\theta)$ in \eqref{M}.

\begin{Proposition} \label{probD}
Let $\xi^*$ denote the locally $\phi_p$-optimal design of the form \eqref{xi}  in the dose-response model \eqref{eq2}
with masses  $   w^*_1 ,  \ldots  ,   w_k^*,$
at the points $d^*_1,  \ldots , d^*_k$, respectively. The design $\tilde \xi^* $ with masses $  \tilde w^*_1= \rho_p({1+\rho_p} )^{-1}  w^*_1 ,  \ldots  ,   \tilde w^*_k = \rho_p({1+\rho_p} )^{-1}  w_k^*,$ and
$   \tilde w^*_{k+1} =  ({1+\rho_p} )^{-1}$
at the points $(d^*_1,0),  \dots , (d^*_k,0)$ and $  (C,1)$, respectively,
is locally $\phi_p$-optimal in the dose-response model  \eqref{eq2} with an active control  \eqref{eq2a}, where
\begin{equation} \label{rho}
\rho_p =
\begin{cases}
 \frac {({\tr}[\{ \I^{-1}_2(\theta_2) \}^{-p}])^{1/(p-1)}}
{({\tr}[\{M^{-1}_1(\tilde \xi^*, \theta_1)\}^{-p}])^{1/(p-1)}} & \mbox{if } p \in (-\infty ,1) \setminus \{0\}  \\
 \frac{s_1}{2} & \mbox{if } p =0  \\
 \frac{\lambda_{{\min}} (\I_2(\theta_2) ) }{\lambda_{{\min}} (M_1(\tilde \xi^*, \theta_1)) }
 & \mbox{if } p =- \infty   \\
\end{cases} ~.
\end{equation}
\end{Proposition}

We note that
Proposition \ref{probD}  can be extended to construct  minimally supported designs. In particular, any minimally supported
$\phi_p$-optimal design  of the form \eqref{xi}  for  the dose response model \eqref{eq2} yields
a minimally supported $\phi_p$-optimal design for the dose response model  \eqref{eq2} with an active control  \eqref{eq2a}
by the transformation described in Proposition \ref{probD}.
We conclude this section by  constructing minimally supported $D$-optimal designs for some of the models considered in Section~\ref{sec32}.

\begin{Example} \label{exefftoxACD}
{\rm Assume that the effect of the drug on  efficacy and toxicity  are both studied using Emax models. The minimally
supported $D$-optimal design for model \eqref{eq2} with an active control \eqref{eq2a} can be obtained from Theorem \ref{thmefftoxminiLIN} part (4b)  and  Proposition \ref{probD}. We set $s_1=6$  and  Theorem
\ref{thmefftoxminiLIN}
provides the support points of the minimally supported $D$-optimal design  for  the dose-response model \eqref{eq2}. Proposition \ref{probD}
yields  $\tilde \omega^*_4 =1/4$ for the proportion of patients
treated with the active control.  Additionally, the minimally supported
 $D$-optimal design for model \eqref{eq2} with an active control \eqref{eq2a} allocates
 the rest of the patients equally to the new drug at $3$ dose levels given by

$$
L, ~\frac{\sqrt{(L+\vartheta_2^e) (L+\vartheta_2^t) (R+\vartheta_2^e) (R+\vartheta_2^t)}+L R-\vartheta_2^e \vartheta_2^t}{L+R+\vartheta_2^e+\vartheta_2^t}
\mbox{  and  } ~R.
$$
In a similar manner explicit results for the other models considered in Section \ref{sec32} can be obtained
 (and are omitted for space considerations). 
 %Further examples can be found in the following section.
}
\end{Example}

%%%%%%%%%------------------------------------%%%%%%%%%%%%%%%%%%%%%%%%%%%%%%%%

\section{Examples}\label{sec5}
\def\theequation{5.\arabic{equation}}
\setcounter{equation}{0}

We now apply our results from previous sections and construct optimal designs for active controlled trials for  three examples.
 In the  first one we determine the locally $D$-optimal design for a particular scenario of the motivating example in  the introduction.
 The second example  
 compares the $D$-optimal design with the $E$-optimal design, which is another type of optimal design sometimes 
 used for making inference on the model parameters. The third
 example contrasts $D$-optimal designs with minimally supported $D$-optimal designs
  with recommendations on their use in practice from a statistical viewpoint.

If the optimal designs are not minimally supported they usually have to be  determined
numerically and several
 algorithms have been proposed in the literature for this purpose.  The optimal designs presented in this section 
 are found
 using particle swarm optimization (PSO), which is a prominent member of the class of  nature-inspired metaheuristic algorithms.
 PSO has been  widely used to solve hard and large dimensional optimization problems in engineering and computer science,
and   it has only been used recently to find optimal designs [see
\cite{kimli11},  \cite{chen2014} or \cite{phoa}].   For space consideration, we omit details  on PSO and refer the interested reader to \cite{qiu}
and  \cite{wongmix} for  details and illustrations.

\begin{Example}  \label{ex51}
{\rm 
\cite{talipi2015} used an  Emax-model
with parameters $\theta_1^e=(2.5,14.5,0.2)^T$ for the mean efficacy outcome
 and  an exponential model 
 $\eta_1^t(d,\theta_1^t)=0.163 + 0.037 e^{(3.3\log(6)d)}$ to model   the toxicity effects
[see Table 1 in  this reference]. As described in Section 3.3.1 of \cite{talipi2015},  they
used a uniform design to   allocate patients to the dose
 levels $0$, $0.05$, $0.2$, $0.4$, $0.6$, $0.8$,  and $1$, respectively. For the  error distribution in model \eqref{eq1}
 they   assumed a two dimensional centered normal distribution with parameters $\rho=0.4$, $\sigma_e=7$ and $\sigma_t=8$.
We simulated data according to model \eqref{eq1} with sample sizes
 $n_1=350$ for the new drug and fitted an Emax and the quadratic model for efficacy and toxicity, respectively.
 The quadratic model was used, because it yields a similar shape as the exponential model and minimally supported
 designs are explicitly available for the combinations of the  Emax and a quadratic model.   The fits  of both 
 regression models to the simulated data are  shown  in Figure \ref{figneu}.
The estimates for the parameters are given by
 $\hat \theta_1^e=(2.588,15.64,0.26)$ and $\hat \theta_1^t=(0.24,-11.632,25.11)$ for the Emax and  quadratic model, while the
 estimates for the covariance are obtained as $\hat \rho=0.387$, $\hat \sigma_e=7.272$ and $\hat \sigma_t=8.311$.
 We used this information to determine a locally $D$-optimal design for the active controlled trial. Note that we do not require information from the model
   for the active control for this purpose as we are calculating $D$-optimal designs [see Proposition \ref{probD}].
 \begin{figure}[t]
\centering
 \includegraphics[width=6.5cm]{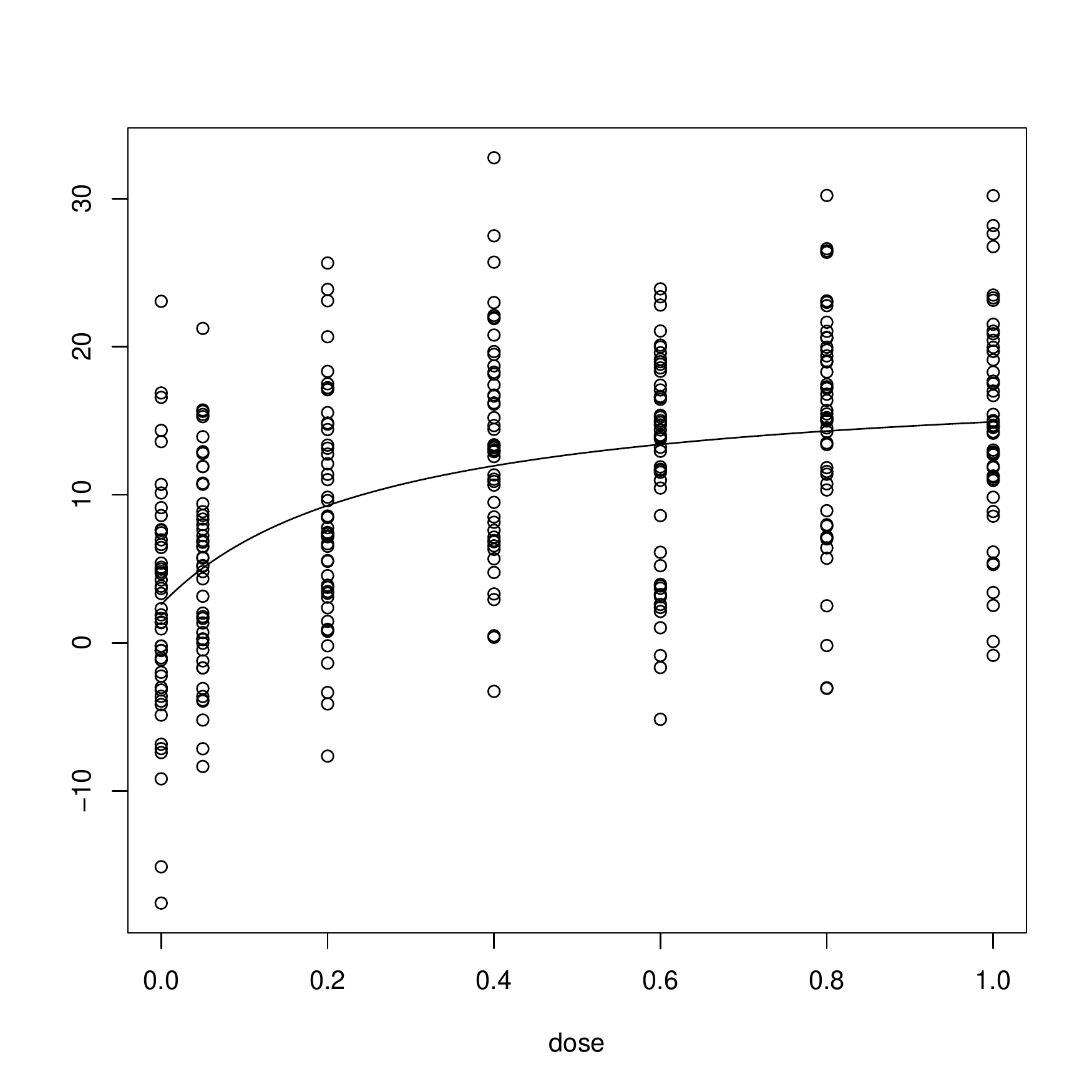}
~~~~~~~~~~
\includegraphics[width=6.5cm]{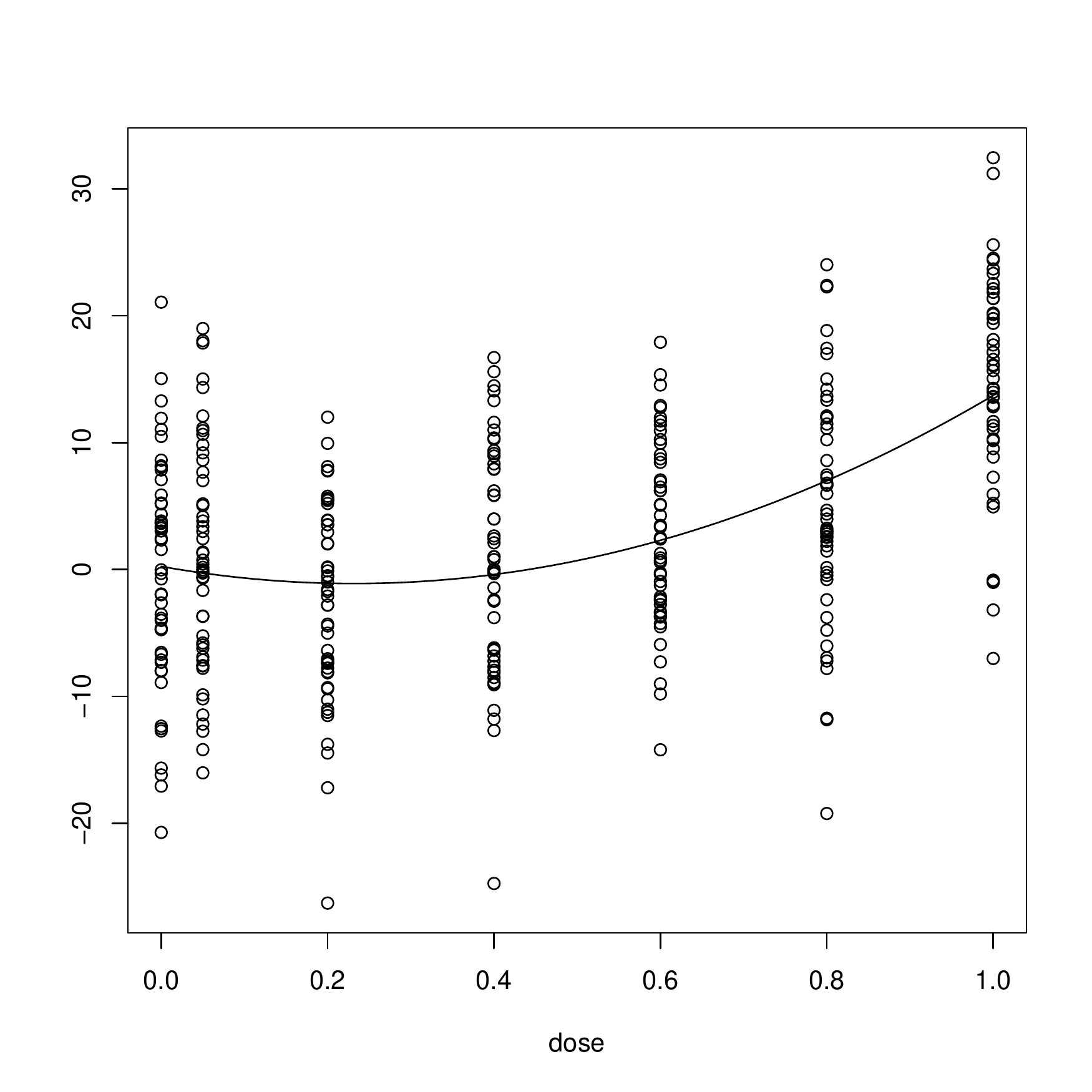}
\caption{\it  Fit of an Emax (efficacy) and a quadratic model to the data generated by a model discussed in  \cite{talipi2015}.}
\label{figneu}
\end{figure}

 By Theorem \ref{thmefftoxQUAD}(c) and Theorem \ref{thethmadm}, we only need to consider
  designs with at most six support points. We first used the PSO algorithm to generate the locally
  $D$-optimal design for model \eqref{eq1} and in the second step, applied Proposition \ref{probD} to determine the locally optimal design
  for the model with an active control. The results are  shown in Table \ref{optdesignsmotivatingexample}. The locally D-optimal design
  has five support points and is therefore not minimally supported. The minimally supported D-optimal design
  can be obtained from Theorem \ref{thmefftoxminiLIN} (4a)  and
  is  shown in the right part of Table \ref{optdesignsmotivatingexample}. The optimality of the design for the new drug was checked by Theorem \ref{equithm}.
 Figure \ref{equivData} displays the sensitivity function of  the locally D-optimal and the minimally supported D-optimal
 design. The results confirm its  optimality and its  non-optimality, respectively.
The $D$-efficiency of the minimally supported designs is given by $0.9886$. We  note that the 
 lower bound \eqref{effbound}   for the $D$-efficiency of the minimally supported optimal design  does not need
the knowledge of the locally $D$-optimal design and is given by $0.9532$.

The good performance of the minimally supported design is also confirmed by calculating the $D$-efficiency of the uniform design 
used in \cite{talipi2015} relative to our locally $D$-optimal design and the minimally supported $ D$-optimal design.  
These relative efficiencies are $0.575$ and $0.581$, respectively, showing that  the performance of the design
implemented by  \cite{talipi2015} could be substantially improved by using locally $D$-optimal designs.

\begin{table}
\begin{center}
\begin{footnotesize}
\begin{tabular}{|c|c|} \hline
$D$-optimal design & minimally supported D-optimal design   \\
\hline
\begin{tabular}{c c c c c}
$(0,0)$ & $(0.18,0)$ & $(0.49,0)$& $(1,0)$ & $(C,1)$\\
\hline
%$0.31$ & $0.18$ & $0.18$ & $0.31$
$0.09$ & $0.16$ & $0.16$ & $0.09$ & $0.5$
\end{tabular}
 &
\begin{tabular}{c c c c}
$(0,0)$ & $(0.31,0)$& $(1,0)$ & $(C,1)$\\
\hline
$0.25$ & $0.25$ & $0.25$ & $0.25$
\end{tabular}
\\
\hline
\end{tabular}
\caption{\it Locally $D$-optimal design and minimally supported D-optimal design for a situation discussed in  \cite{talipi2015}.
The efficacy is modeled by an Emax  model and the toxicity by a quadratic model.}
\label{optdesignsmotivatingexample}
\end{footnotesize}
\end{center}
\end{table}

\begin{figure}[h]
\centering
     \subfigure{\includegraphics[width=6.cm]{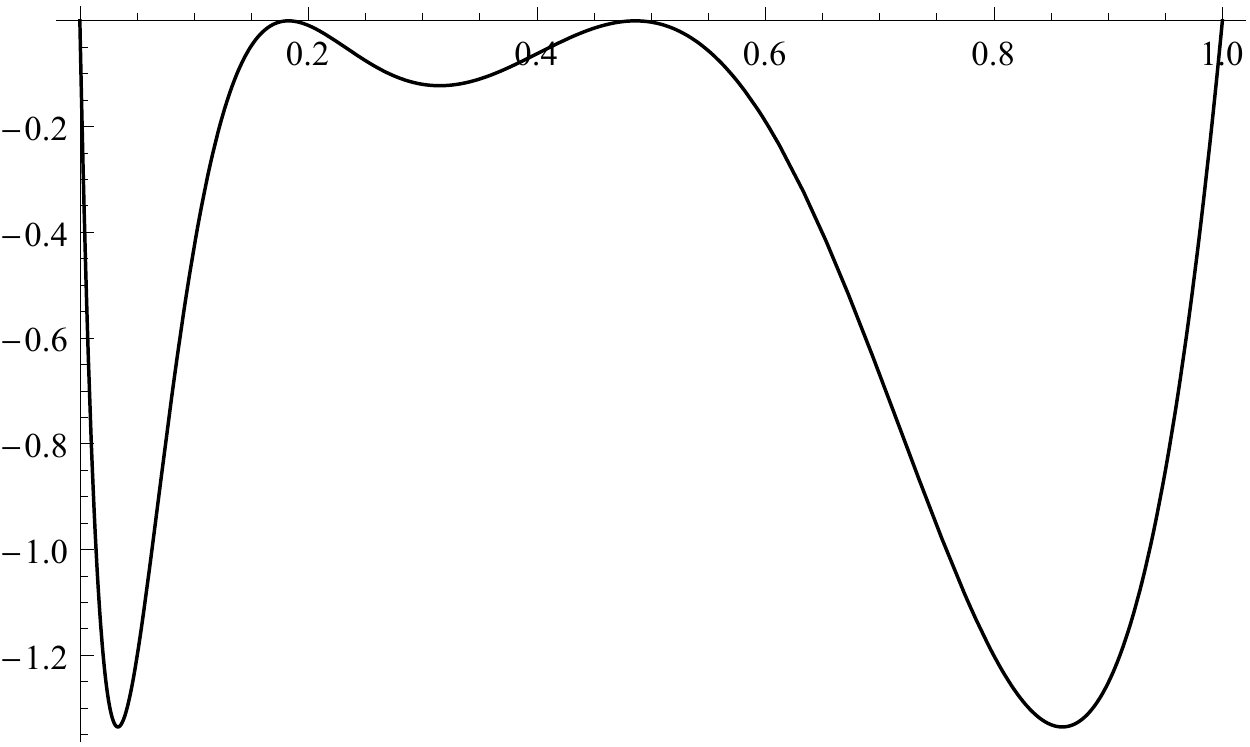}}
~~~~~~~~~~~~~
\subfigure{\includegraphics[width=6.3cm]{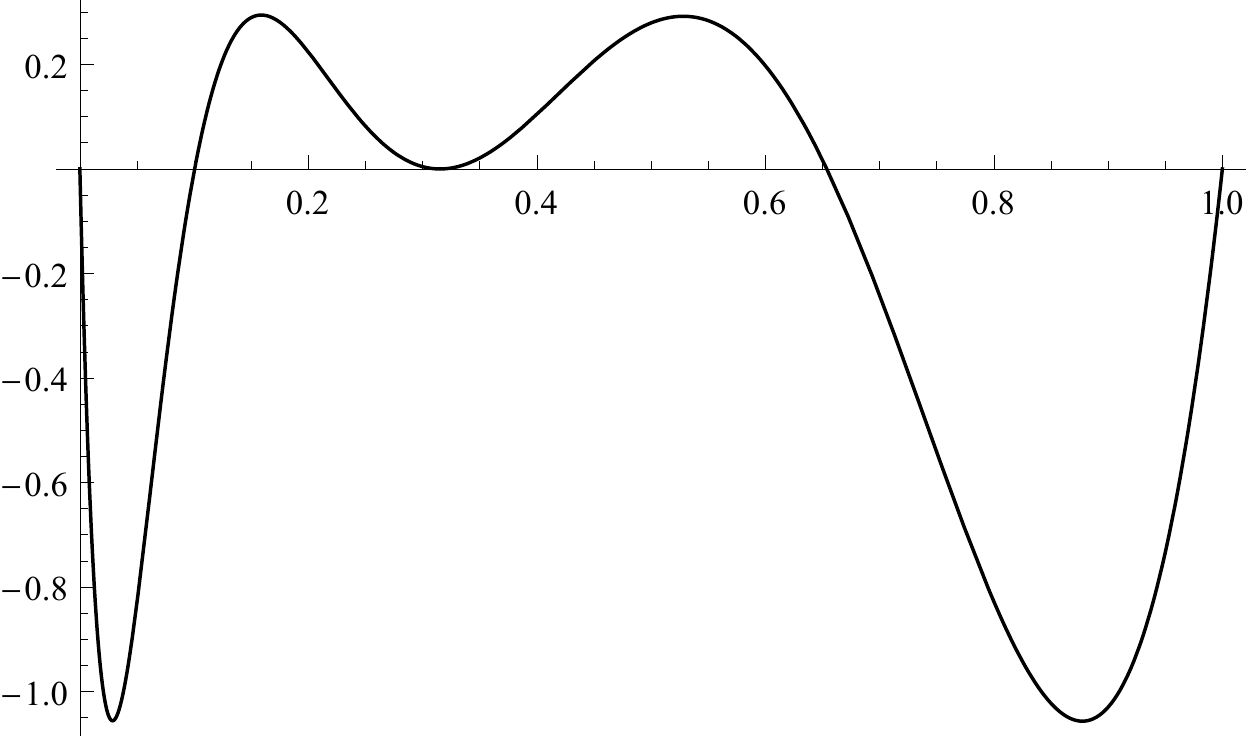}}
\caption{\it The sensitivity functions of the locally $D$-optimal design and the minimally supported D-optimal design for an active-controlled trial. The designs are given in   Table \ref{optdesignsmotivatingexample}. The  efficacy is modeled  by an Emax  model and   the toxicity by a quadratic model.}
\label{equivData}
\end{figure}
}
\end{Example}

\begin{table}[t]
\begin{center}
\begin{footnotesize}
\begin{tabular}{|c|c|c|} \hline
$\rho$ & $D$-optimal design   & $E$-optimal design   \\
\hline
 $0.1$ & \begin{tabular}{c c c c}
$(0,0)$ & $(23.84,0)$ & $(150,0)$ & $(C,1)$ \\
\hline
$0.16$ & $0.31$ & $0.31$ & $0.22$ \\
\end{tabular}  &\begin{tabular}{c c c c}
$(0,0)$ & $(19.08,0)$ & $(150,0)$ & $(C,1)$ \\
\hline
$0.22$ & $0.47$ & $0.25$ & $0.06$ \\
\end{tabular}   \\
\hline
$0.5$ & \begin{tabular}{c c c c}
$(0,0)$ & $(23.84,0)$ & $(150,0)$ & $(C,1)$ \\
\hline
$0.16$ & $0.31$ & $0.31$ & $0.22$ \\
\end{tabular}   & \begin{tabular}{c c c c}
$(0,0)$ & $(19.37,0)$ & $(150,0)$ & $(C,1)$ \\
\hline
$0.15$ & $0.49$ & $0.31$ & $0.05$  \\
\end{tabular}   \\
\hline
$0.8$ & \begin{tabular}{c c c c}
$(0,0)$ & $(23.84,0)$ & $(150,0)$ & $(C,1)$ \\
\hline
$0.16$ & $0.31$ & $0.31$ & $0.22$ \\
\end{tabular}   & \begin{tabular}{c c c c}
$(0,0)$ & $(18.65,0)$ & $(150,0)$ & $(C,1)$ \\
\hline
$0.11$ & $0.51$ & $0.33$ & $0.05$  \\
\end{tabular}  \\
\hline
\end{tabular}
\caption{\it Locally
$D$- and $E$-optimal designs for an active-controlled trial, where   the efficacy is modeled by an Emax  
model
and  the toxicity by a Michaelis-Menten model.  The parameters in the two models are
 $\theta_1^e = (0, 0.466, 25)^T, \theta_1^t = (300,50)^T, \sigma_e=0.2, \sigma_t=20 ,\sigma_e^{AC}=0.2, \sigma_t^{AC}=29.8$ and $\rho \in \{0.1, 0.5, 0.8\}$.}
\label{optdesignsexample}
\end{footnotesize}
\end{center}
\end{table}

\begin{figure}[h]
\centering
 	\subfigure{\includegraphics[width=6.cm]{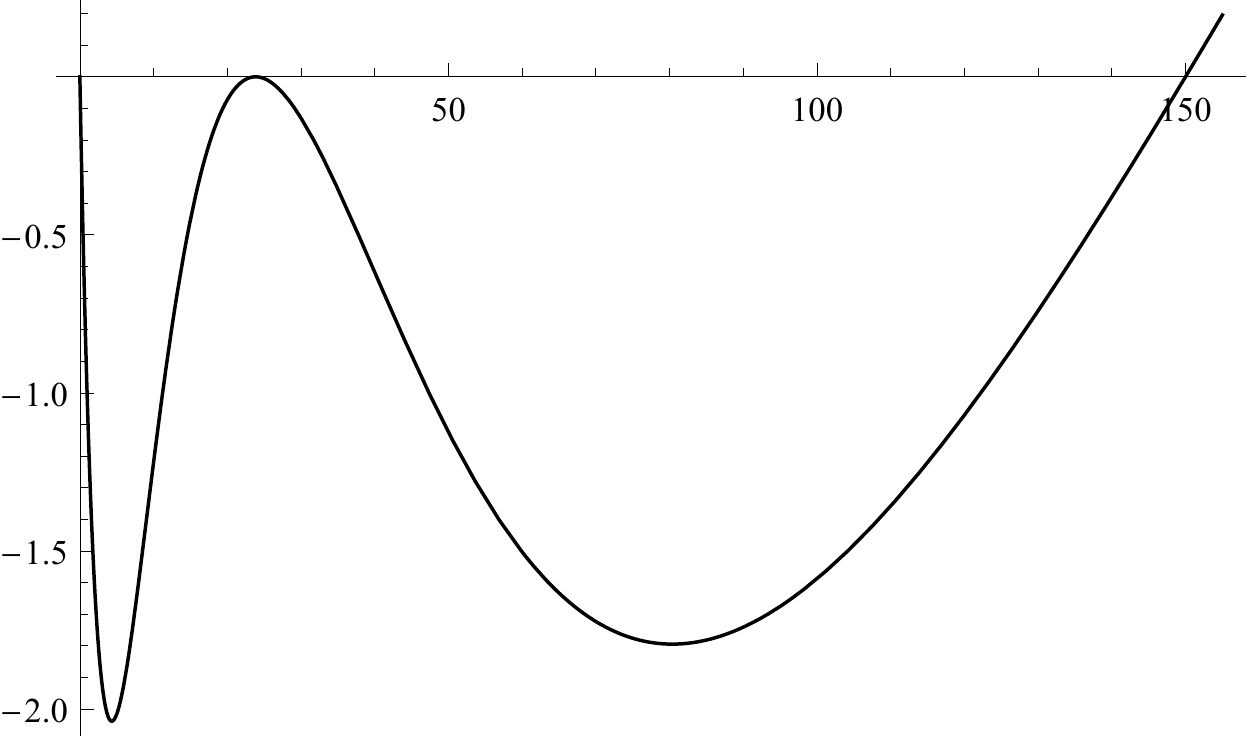}}
~~~~~~~~~~~~~
 	\subfigure{\includegraphics[width=6.3cm]{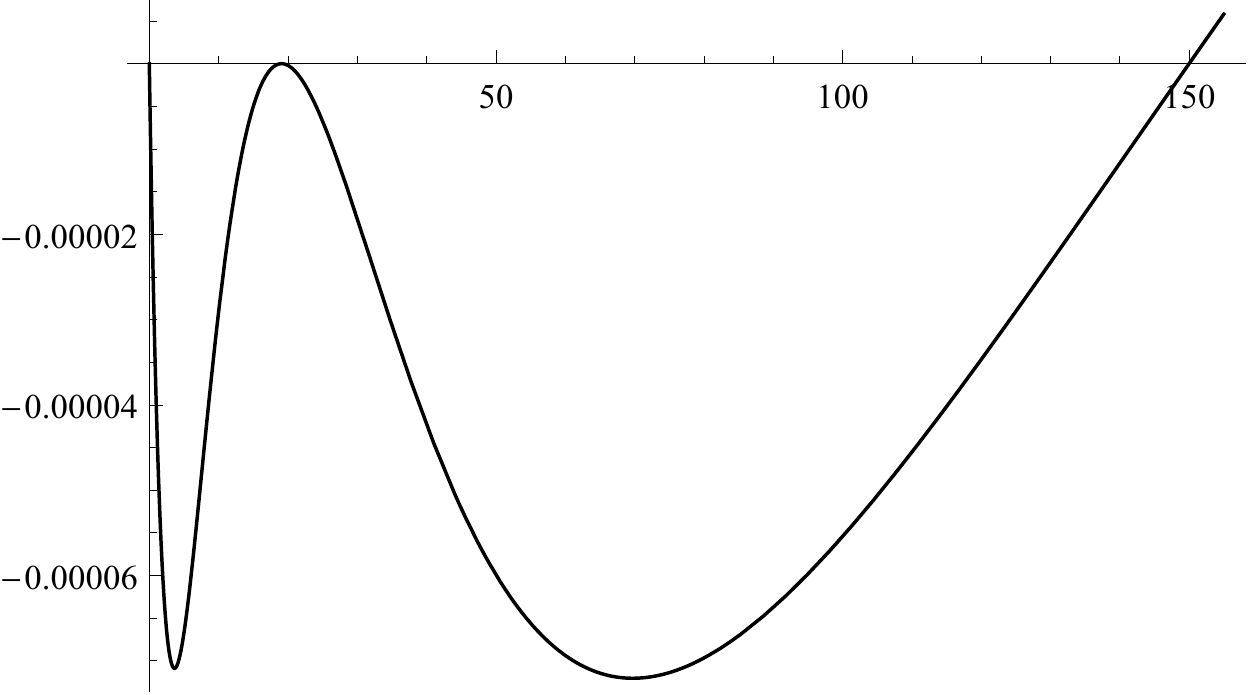}}
\caption{\it The sensitivity function of the locally $D$-optimal design (left) and the locally $E$-optimal design (right)
confirm the optimality of the PSO-generated designs. The   efficacy is modeled  by an Emax  and   the toxicity is modeled by a 
Michaelis-Menten model, where the correlation between efficacy and toxicity is given by $\rho =  0.1$.}
\label{equivD}
\end{figure}

\begin{Example} \label{ex52}
{\rm Consider a situation where the efficacy outcome is described by an Emax model and a Michaelis-Menten model is used for the toxicity outcome. The nominal parameter values are  $\theta_1=(0, 0.466, 25, 300, 50)^T$, and   the dose interval is  $\mathcal{D}=[0,150]$. We chose $\sigma_e=0.2$,
 $\sigma_t=20$ and various values for the correlation in the covariance matrix are considered. By Theorem \ref{thmefftoxMM}(b) and Theorem \ref{thethmadm}, we only need to consider designs with at most six support points.  We applied the PSO algorithm 
 to generate the locally $D$- and $E$-optimal designs for model \eqref{eq1} and  Proposition \ref{probD} 
  to determine the locally optimal designs for the dose finding trial  with an active control. The results are shown in Table \ref{optdesignsexample} for various values of the   correlation $\rho$. By definition, an $E$-optimal design minimizes the maximum   eigenvalue of the inverse of the information matrix,  whereas a $D$-optimal design  minimizes the volume of the confidence ellipsoid for the   parameter.  The locally $D$- and $E$-optimal designs for the active controlled trial have four support points and are therefore minimally supported. Consequently, the support points of the $D$-optimal designs do not depend on the elements of the covariance matrix $\Sigma_1$, as predicted by Theorem \ref{equallyweighted}. On the other hand, the interior support points of the $E$-optimal design are slightly changing with the correlation $\rho$.  The optimality of both designs was checked by Theorem \ref{equithm} and Figure \ref{equivD}
%and \ref{equivE}
displays the sensitivity functions of the designs   that confirm their optimality for $\rho=0.1$.
}
\end{Example}

\begin{figure}
\centering
  \subfigure{\includegraphics[width=5.3cm]{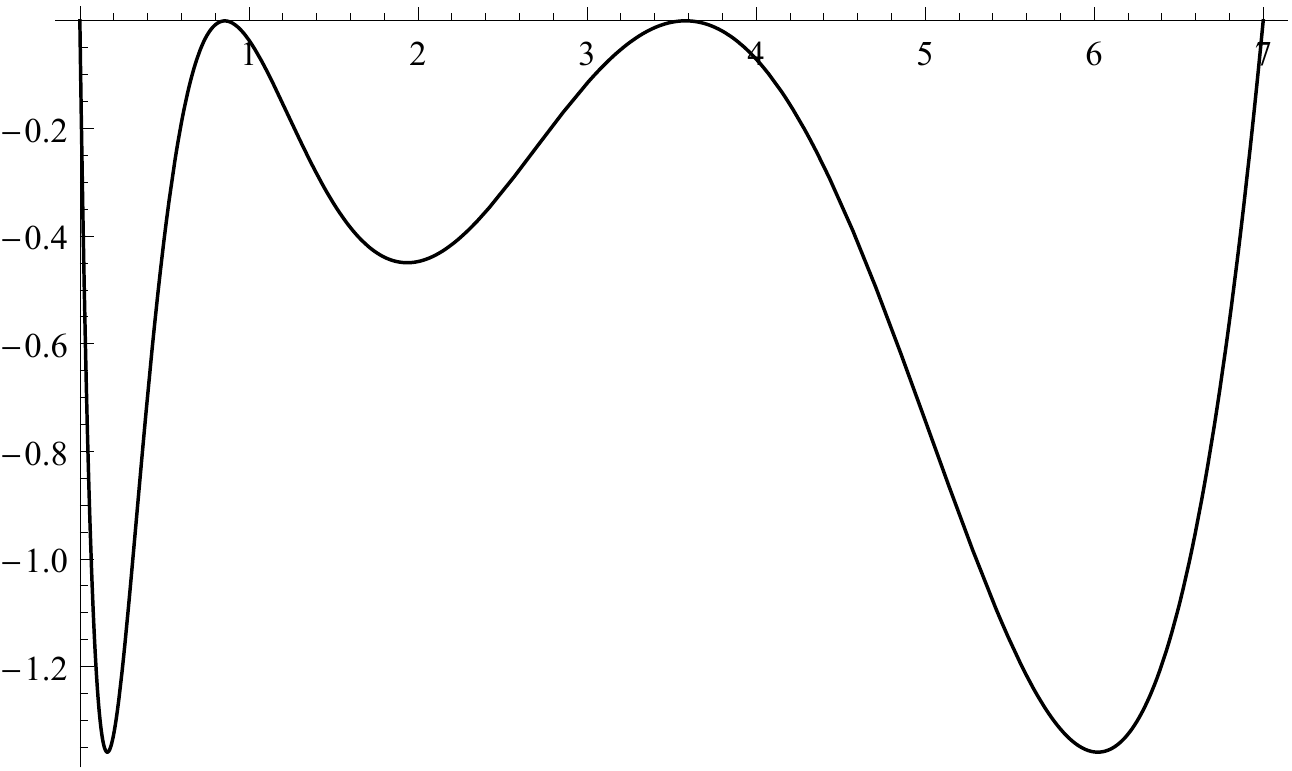}}
	\hfill
 	\subfigure{\includegraphics[width=5.3cm]{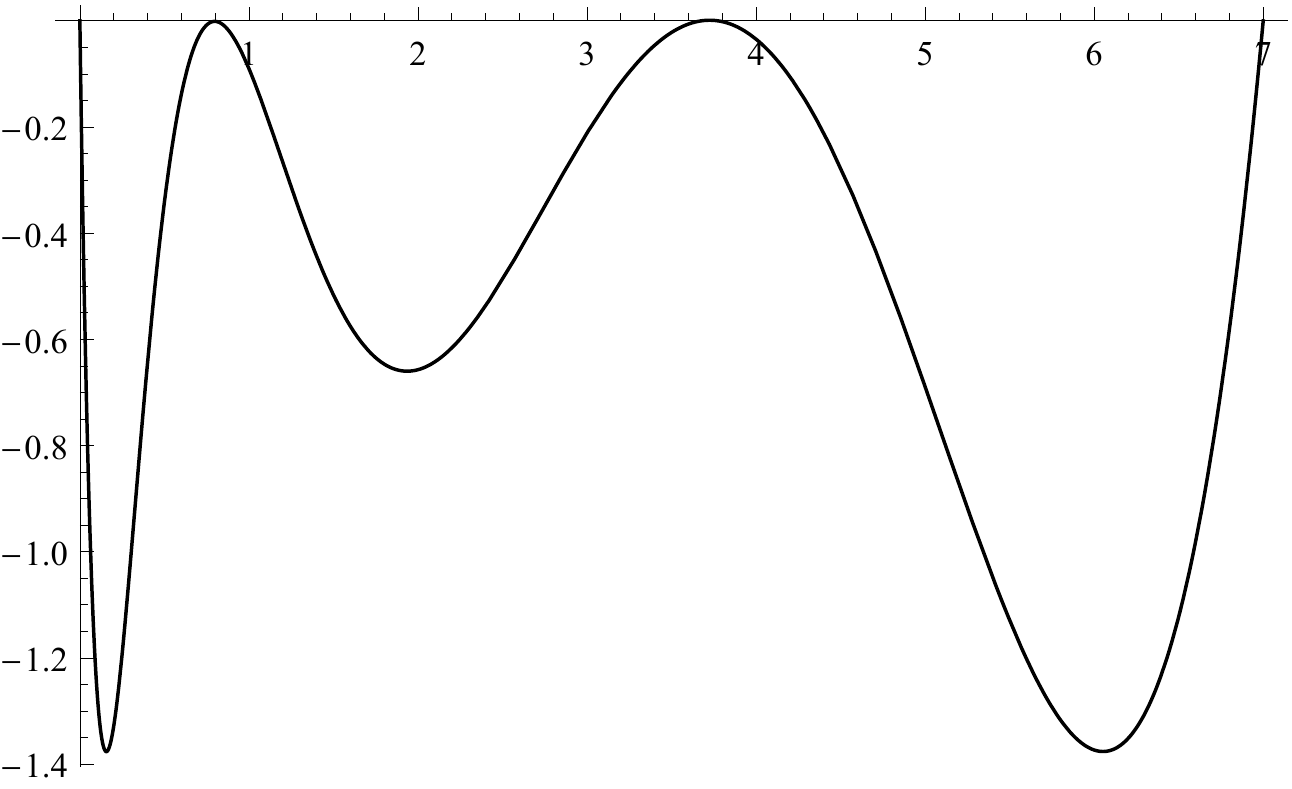}}
		\hfill
 	\subfigure{\includegraphics[width=5.3cm]{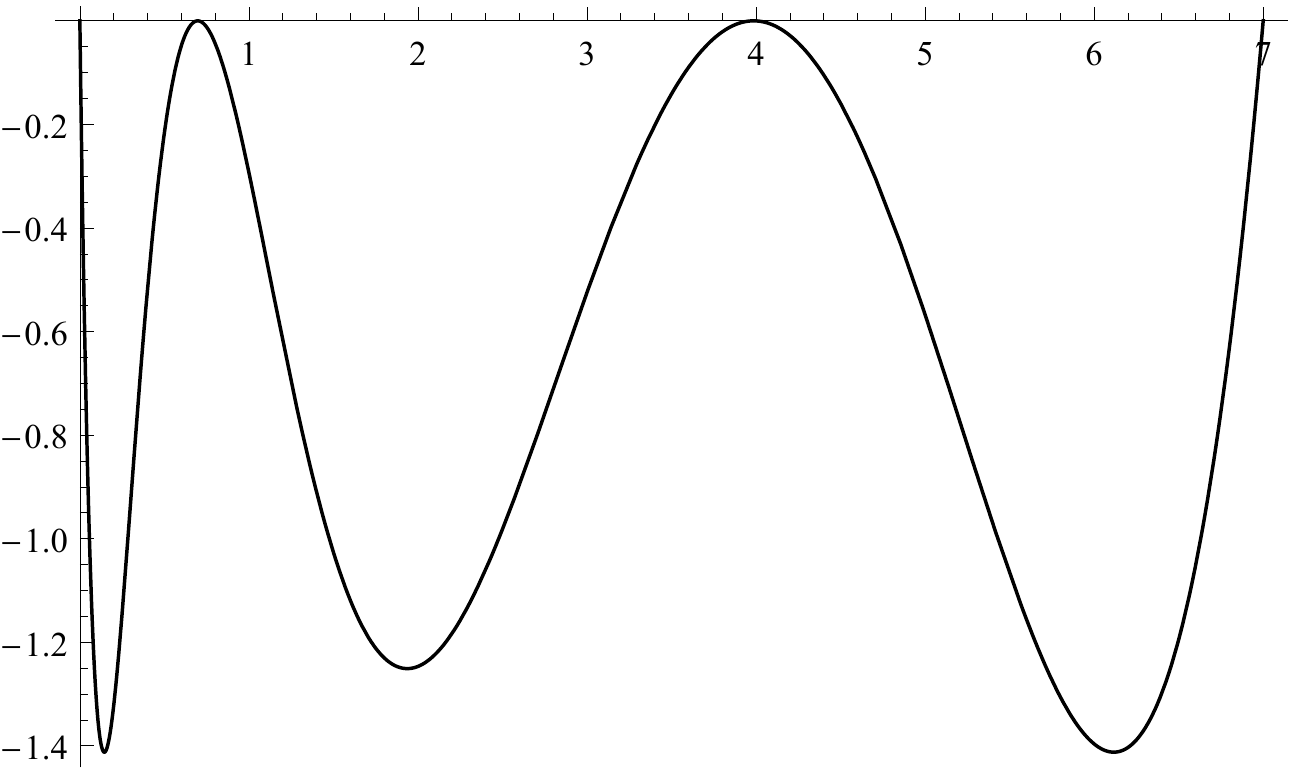}}
\caption{\it The sensitivity functions of the  locally $D$-optimal design in Table \ref{optdesignsexamplezwei}
 confirm their optimality. 
%The mean outcome is represented by the quadratic and Emax-model with $\rho = 0.1$(left), $0.5$(middle) and $0.9$(right). 
}
\label{equiv}
\end{figure}

\begin{Example} \label{ex53}
{\rm Assume that   the efficacy outcome is described by a quadratic model and the toxicity outcome by an Emax-model, where  the nominal values of the model parameters are given by $\theta_1=(0.5, 0.01, 0.1,0.1, 2.4, 1.2)^T$.  The dose interval is $\mathcal{D}=[0,7]$ and we chose $\sigma_e=0.1$ and $\sigma_t=0.4$. It follows from Theorem \ref{thmefftoxQUAD}(d) and Theorem \ref{exefftoxAC}   that   only     designs with at most six support points have to be considered.      The locally $D$-optimal designs are determined in the same way as described in Example \ref{ex51} and \ref{ex52} and the   results are listed in the left part of Table \ref{optdesignsexamplezwei} for different values of the correlation. Note that the $D$-optimal designs are not minimally supported and the support points and weights  depend on   the correlation.  The minimally supported   $D$-optimal designs can be found by an application of  Theorem \ref{thmefftoxminiLIN}   and do not depend on $\rho$ [see the right part of Table \ref{optdesignsexamplezwei}].
The optimality of the numerically calculated $D$-optimal designs was checked by Theorem \ref{equithm}  and the corresponding  sensitivity functions are displayed in Figure \ref{equiv} for different values of the correlation, that is   $\rho=0.1, 0.5$ and $0.9$. We observe that all designs calculated by the metaheuristic PSO-algorithm are in fact $D$-optimal. Moreover, the efficiencies of the minimally supported designs are given by $0.96, 0.81$ and $0.34$ for the case $\rho = 0.1, 0.5,$ and $0.9$, respectively. 
%The sensitivity functions corresponding to the minimally supported $D$-optimal designs are shown in Figure \ref{notequiv} and it is obvious that they are not locally $D$-optimal.
From the efficiencies we see that the minimally supported designs are only efficient if the efficacy and toxicity outcomes are nearly uncorrelated. For a strong correlation between efficacy and toxicity minimally supported designs cannot be recommended. Finally, we note that the
 values of the lower bounds
 in \eqref{effbound} for these $3$ minimally supported optimal designs are  $0.87, 0.67$ and $0.18$.}
\end{Example}

\begin{table}
\begin{center}
\begin{footnotesize}
\begin{tabular}{|c|c|c|} \hline
$\rho$ & optimal   & minimally supported $D$-optimal   \\
\hline
$0.1$ & \begin{tabular}{c c c c c}
$(0,0)$ & $(0.86,0)$& $(3.58,0)$ & $(7,0)$ & $(C,1)$\\
\hline
$0.225$ & $0.15$ & $0.15$ & $0.225$ & $0.25$ \\
\end{tabular}   & \begin{tabular}{c c c c}
$(0,0)$ & $(1.94,0)$ & $(7,0)$  & $(C,1)$\\
\hline
$0.25$ & $0.25$  & $0.25$ & $0.25$ \\
\end{tabular}   \\
\hline
$0.5$ & \begin{tabular}{c c c c c}
$(0,0)$ & $(0.8,0)$ & $(3.73,0)$ & $(7,0)$ & $(C,1)$\\
\hline
$0.2175$ & $0.1575$ & $0.1575$ & $0.2175$ & $0.25$ \\
\end{tabular}   & \begin{tabular}{c c c c}
$(0,0)$ & $(1.94,0)$ & $(7,0)$  & $(C,1)$ \\
\hline
$0.25$ & $0.25$  & $0.25$ & $0.25$ \\
\end{tabular}   \\
\hline
$0.9$ & \begin{tabular}{c c c c c}
$(0,0)$ & $(0.7,0)$& $(3.99,0)$ & $(7,0)$ & $(C,1)$\\
\hline
$0.21$ & $0.165$ & $0.165$ & $0.21$ & $0.25$ \\
\end{tabular}    & \begin{tabular}{c c c c}
$(0,0)$ & $(1.94,0)$ & $(7,0)$  & $(C,1)$\\
\hline
$0.25$ & $0.25$  & $0.25$& $0.25$ \\
\end{tabular}   \\
\hline
\end{tabular}

\caption{\it Locally $D$-optimal design (left) and the minimally supported $D$-optimal  designs (right).  The efficacy and toxicity are modeled  by a quadratic model with parameter $\theta_1^e=(0.5, 0.01, 0.1)^T$ and Emax-model with parameter $\theta_1^t = (0.1, 2.4, 1.2)^T$, respectively. The elements in the covariance matrix are $\sigma_e=0.1, \sigma_t= 0.4$ and various correlation values.}
\label{optdesignsexamplezwei}
\end{footnotesize}
\end{center}
\end{table}

%\begin{figure}
%\centering
%  \subfigure{\includegraphics[width=5.3cm]{EMAX_QUAD_mini_equiv_rho01.pdf}}
%	\hfill
% 	\subfigure{\includegraphics[width=5.3cm]{EMAX_QUAD_mini_equiv_rho05.pdf}}
%		\hfill
% 	\subfigure{\includegraphics[width=5.3cm]{EMAX_QUAD_mini_equiv_rho09.pdf}}
%\caption{\it The sensitivity function for the minimally supported $D$-optimal design displayed in the right part of Table \ref{optdesignsexamplezwei}. The efficacy and toxicity are modeled by a  quadratic and an Emax-model, respectively, and different correlations are considered: $\rho = 0.1$(left), $0.5$(middle) and $0.9$(right). }
%\label{notequiv}
%\end{figure}

 \section{Conclusions and further research} \label{sec6}
 \def\theequation{6.\arabic{equation}}
 \setcounter{equation}{0}
 
 In this paper  we investigated the optimal design problem  for active controlled  trials with bivariate outcomes. Upper bounds on 
 the number of support points of  locally optimal have been derived, which are used to
 reduce the dimensionality of the corresponding optimization problems.  We also determined minimally  supported $D$-optimal
 designs  explicitly for specific combinations of models for the efficacy and toxicity and note that in
 general the optimal designs for active controlled clinical trials with bivariate outcomes are not minimally supported.
 Nevertheless, it is demonstrated that for the models under consideration 
  the minimally supported $D$-optimal designs are rather efficient, provided that the correlation between efficacy and toxicity 
  is weak. Our results demonstrate that  statistical inference in clinical trials with bivariate outcomes  can be improved
  substantially by the appropriate use of efficient  designs. 
  
This paper discusses  locally optimal designs, which require a-priori information about the unknown model parameters if they 
appear in  the  model in  a nonlinear  way
[see \cite{chernoff1953}].  When preliminary knowledge regarding the unknown parameters of a nonlinear model is available, and 
the application of locally optimal designs is well justified [see for example \cite{debrpepi2008}]. 
Locally optimal  designs  are typically  used  as benchmarks for commonly used 
designs [see the discussion in Example \ref{ex51}]. 
Additionally, locally optimal designs serve as basis for constructing optimal  designs with respect to more sophisticated optimality
criteria,  which are robust against a misspecification of the unknown parameters;
see \cite{pronwalt1985}  or \cite{chaver1995}, \cite{dette1997} among others.  An interesting direction for future research 
is to further develop  the methodology introduced in the present paper
to address uncertainty in the preliminary information for the unknown parameters.

 \bigskip 
 \bigskip

{\bf Acknowledgements} The authors would like to thank Martina
Stein, who typed parts of this manuscript with considerable
technical expertise.
This work has been supported in part by the
Collaborative Research Center ``Statistical modeling of nonlinear
dynamic processes'' (SFB 823) of the German Research Foundation
(DFG).  Kettelhake, Schorning  and Wong were partially supported  by a grant from the National Institute Of General Medical Sciences of the National
Institutes of Health under Award Number R01GM107639. The content is solely the responsibility of the authors and does not necessarily
 represent the official views of the National
Institutes of Health.
%The authors would also like to thank
%  two anonymous referees  and the associate editor for very constructive comments on an earlier version of this paper.

%%%%%%%%%------------------------------------%%%%%%%%%%%%%%%%%%%%%%%%%%%%%%%%

\section{Appendix} \label{sec7}
\def\theequation{7.\arabic{equation}}
\setcounter{equation}{0}

% \subsection {Preliminaries}  \label{prel}

\subsection {Proof of Theorems \ref{thmefftoxLIN}, \ref{thmefftoxQUAD}, \ref{thmefftoxMM} and \ref{thmefftoxEMAX}}
 \label{sec71}

We present the proof of Theorem \ref{thmefftoxEMAX} only for the case, where  the effect of the drug on efficacy and toxicity is  modeled by an Emax model.
In this case   the gradient of the outcome with respect to the parameter is given by
\begin{eqnarray*}
	 \frac{\partial}{\partial \theta_1}\eta_1(d,\theta_1) &=& \begin{pmatrix} 1
		& \frac{d}{\vartheta_2^e+d} & -\frac{\vartheta_1^e d}{(\vartheta_2^e+d)^2} & 0 & 0 & 0 \\ 0 & 0 & 0 & 1 & \frac{d}{\vartheta_2^t+d} & -\frac{\vartheta_1^t d}{(\vartheta_2^t+d)^2}
		 \end{pmatrix}.
%		 \\
%		&=& \begin{pmatrix} 1 & 1-\frac{\vartheta_2^e}{\vartheta_2^e+d} & \frac{\vartheta_1^e \vartheta_2^e }{(\vartheta_2^e+d)^2}-\frac{\vartheta_1^e}{\vartheta_2^e+d} & 0 & 0 & 0 \\ 0 & 0 & 0 & 1 & %1-\frac{\vartheta_2^t}{\vartheta_2^t+d} & \frac{\vartheta_1^t \vartheta_2^t }{(\vartheta_2^t+d)^2}-\frac{\vartheta_1^t}{\vartheta_2^t+d}   \end{pmatrix}.
	\end{eqnarray*}
It is easy to see that there exists a full column rank  matrix $L \in \R^{6 \times 10} $ which does not depend on the variable $d$ such that
$$
 \frac{\partial}{\partial \theta_1}\eta_1(d,\theta_1)   =  \begin{pmatrix}
  \nu^T(d) &0   \\  0  & \nu^T(d) \end{pmatrix}  L^T~,
$$
where the vector  $\nu (d)$ is defined by the linearly independent functions in the gradient, i.e.
$$
\nu(d)=(1,\tfrac{1}{\vartheta_2^e+d},\tfrac{1}{(\vartheta_2^e+d)^2},\tfrac{1}{\vartheta_2^t+d},\tfrac{1}{(\vartheta_2^t+d)^2})^T \in \R^5.
$$
Consequently, we obtain for the information matrix in \eqref{Mone}
the representation
	\begin{equation} \label{MLSigma}
		M_1(\xi, \theta_1) =  L \int_{\D} \begin{pmatrix} \nu(d)\nu^T(d) & \mathbf{0}  \\ \mathbf{0}  & \nu(d)\nu^T(d) \end{pmatrix} d \xi(d)  L^T,
	\end{equation}
where the matrix $\mathbf{0} $ denotes a $ 5 \times 5$ square matrix with all entries $0$ and
\begin{equation}\label{matmat}
		\nu(d) \nu^T(d) = \begin{pmatrix}
		1 & \tfrac{1}{\vartheta_2^e+d} & \tfrac{1}{(\vartheta_2^e+d)^2} &  \tfrac{1}{\vartheta_2^t+d} &  \tfrac{1}{(\vartheta_2^t+d)^2} \\
		\tfrac{1}{\vartheta_2^e+d} & \tfrac{1}{(\vartheta_2^e+d)^2} & \tfrac{1}{(\vartheta_2^e+d)^3} &   \tfrac{1}{(\vartheta_2^e+d)(\vartheta_2^t+d)}  &  \tfrac{1}{(\vartheta_2^e+d)(\vartheta_2^t+d)^2} \\
		\tfrac{1}{(\vartheta_2^e+d)^2} & \tfrac{1}{(\vartheta_2^e+d)^3} & \tfrac{1}{(\vartheta_2^e+d)^4} & \tfrac{1}{(\vartheta_2^e+d)^2(\vartheta_2^t+d)}  & \tfrac{1}{(\vartheta_2^e+d)^2(\vartheta_2^t+d)^2} \\
		\tfrac{1}{\vartheta_2^t+d} & \tfrac{1}{(\vartheta_2^e+d)(\vartheta_2^t+d)} &\tfrac{1}{(\vartheta_2^e+d)^2(\vartheta_2^t+d)} & \tfrac{1}{(\vartheta_2^t+d)^2} & \tfrac{1}{(\vartheta_2^t+d)^3} \\
		\tfrac{1}{(\vartheta_2^t+d)^2} & \tfrac{1}{(\vartheta_2^e+d)(\vartheta_2^t+d)^2} & \tfrac{1}{(\vartheta_2^e+d)^2(\vartheta_2^t+d)^2} & \tfrac{1}{(\vartheta_2^t+d)^3} & \tfrac{1}{(\vartheta_2^t+d)^4} \end{pmatrix}.
\end{equation}	
Now Theorem 14.2.9 in \cite{harv1997} shows that an improvement with respect to the Loewner ordering can be obtained by improving the common block
$$
 \int_{\D}  \nu(d)\nu^T(d)d \xi(d)
$$
in the matrix \eqref{MLSigma}. For this purpose we now use Theorem \ref{detmelmod}. The functions $\psi_0(d)=1$ and
%The linearly independent  and non-constant
%functions in the matrix  $\nu(d)\nu^T(d)$ are given by
\begin{eqnarray*}
 \psi_1(d) &=&\tfrac{1}{\vartheta_2^e+d}, ~\psi_2(d) = \tfrac{1}{(\vartheta_2^e+d)^2},~ \psi_3(d) = \tfrac{1}{(\vartheta_2^e+d)^3},~ \psi_4(d) = \tfrac{1}{(\vartheta_2^e+d)^4} , \\
\psi_5(d) &=& \tfrac{1}{\vartheta_2^t+d}, ~\psi_6(d) = \tfrac{1}{(\vartheta_2^t+d)^2}, ~\psi_7(d) = \tfrac{1}{(\vartheta_2^t+d)^3}, \psi_8(d) = \tfrac{1}{(\vartheta_2^t+d)^4}.
\end{eqnarray*}
fulfill the conditions specified in the paragraph before Theorem \ref{detmelmod}. It follows by an application of Theorem 1.1 in Chapter IX of \cite{karstud1966} that
  the sets  $\{\psi_0 , \ldots , \psi_7\}$  and $\{\psi_0 , \ldots , \psi_8\}$  are Tchebycheff systems and Theorem
 \ref{detmelmod} is applicable with $k=8$. Part (A2) of this result yields that
 there exists a design $\xi^*$ with at most five support points including $L$ and $R$ such that
  $$ \int_{\D}  \nu(d)\nu^T(d)d \xi(d) \leq  \int_{\D}  \nu(d)\nu^T(d)d \xi^*(d),$$
 and the assertion follows.  We note that an application of Theorem $3.1$ in \cite{detmel2011} is not possible because the different functions from matrix \eqref{matmat} do not form a Tchebycheff system.

\hfill $\Box$

\subsection{Proof of Theorem \ref{equallyweighted}}

Let $\xi$ be a minimally supported design of the form \eqref{xi}. As $s^e_1 = s^t_1$ we have $k=s^t_1+1$. Considering the Cholesky decomposition  $\Sigma^{-1}_1 = \tilde \Sigma \tilde \Sigma^T$ of the inverse of the covariance matrix $\Sigma_1$
we obtain for the information matrix $M_1(\xi,\theta_1)$ the representation
\begin{eqnarray}
	M_1(\xi,\theta_1) \nonumber
		&=& \sum_{i=1}^k \omega_i (\tfrac{\partial}{\partial \theta_1} \eta_1(d_i,\theta_1) )^T \tilde\Sigma \tilde \Sigma^T ( \tfrac{\partial}{\partial \theta_1} \eta_1(d_i,\theta_1)) \\
	&=& G^T 		\mbox{Diag} (\omega_1, \omega_1,\ldots \omega_k, \omega_k)   G, \label{quadmat}
\end{eqnarray}
where the matrix $G$ is defined by
\begin{eqnarray} \label{quadmat1}
G=  \begin{pmatrix} \tilde\Sigma^T (\tfrac{\partial}{\partial \theta_1} \eta_1(d_1,\theta_1) )  \\ \vdots \\ \tilde\Sigma^T (\tfrac{\partial}{\partial \theta_1} \eta_1(d_k,\theta_1) )   \end{pmatrix} ~=~ (I_k \otimes  \tilde\Sigma^T  ) \begin{pmatrix}  (\tfrac{\partial}{\partial \theta_1} \eta_1(d_1,\theta_1) )  \\ \vdots \\
 (\tfrac{\partial}{\partial \theta_1} \eta_1(d_k,\theta_1) )   \end{pmatrix}~\in ~\R^{2k\times 2k} .
\end{eqnarray}
and $A \otimes B$ denotes the Kroecker product of the matrices $A$ and $ B$. Now
$$
\det (M_1(\xi,\theta_1) ) = (\det G)^2 \prod^k_{i=1} w^2_i
$$
and consequently, the minimally supported $D$-optimal design must have equal weights. Moreover, the representation
$$
\det (G)  =  \big( \det ( \tilde\Sigma ) \big) ^k   \det \Big( \big (\tfrac{\partial}{\partial \theta_1} \eta_1(d_j,\theta_1) \big)_{j=1,\ldots ,k}
\big)
$$
shows that the support points of  the minimally supported $D$-optimal design do not depend on the
elements of the matrix $\Sigma_1$. This completes the proof of Theorem \ref{equallyweighted}.
\hfill $\Box$

\subsection{Proof of Theorem  \ref{thmefftoxminiLIN}}

We show only the proof of part 1(b) as the proofs for other cases are similar. 
If a linear and a Michaelis-Menten model are used to describe the effect of the drug on efficacy and toxicity, at least two support points, say $d_1, d_2$, are necessary to guarantee invertibility of the information matrix. From Theorem \ref{equallyweighted} it follows $\omega_1^*=\omega_2^*=\tfrac{1}{2}$. Consider now the determinant of the information matrix
of a design $\xi$ with equal weights at the points $d_1$ and $d_2$, then it follows by a straightforward calculation
that
\begin{equation}\label{det}
	\det(M_1 (\xi,\theta_1)) = \frac{{\vartheta_1^t}^2 d_1^2 d_2^2 (d_1-d_2)^4}{16 \left(\rho^2-1\right)^2 \sigma_e^4 \sigma_t^4 (\vartheta_2^t+d_1)^4 (\vartheta_2^t+d_2)^4}.
\end{equation}
If we assume w.l.o.g. that $d_1 < d_2$, then the right hand side of \eqref{det}  is a monotone function of $d_2$. Consequently
the right boundary point $R$ is one of the optimal support points, that is $d_2=R$. Maximizing the remaining expression with respect to
the point $d_1$ in the interval $[L,R] $ gives
$$
d_1=
L \lor \frac{1}{2}(\sqrt{R^2+10 R \vartheta_2^t+9 (\vartheta_2^t)^2}-R-3 \vartheta_2^t),
$$
which proves the result.  \hfill $\Box$

\subsection{Proof of Theorem \ref{thethmadm}}

Assume that $\tilde \xi$ is not admissible, that is there exists a design
\begin{equation*}
	\tilde \eta = \begin{pmatrix} (\overline{d}_1,0) & \ldots & (\overline{d}_l,0) & (C,1) \\ \overline{\omega}_1 & \ldots & \overline{\omega}_l & \overline{\omega}_{l+1} \end{pmatrix}
\end{equation*}
such that $M(\tilde \eta,\theta_1) \neq M(\tilde\xi,\theta_1)$ and $M(\tilde \eta,\theta_1) \geq M(\tilde\xi,\theta_1)$. This yields immediately $\overline{\omega}_{l+1} \geq \tilde\omega_{k+1}$ and
\begin{equation*}
	(1-\overline{\omega}_{l+1}) M_1(\eta,\theta_1) \geq (1-\tilde{\omega}_{k+1}) M_1(\xi,\theta_1),
\end{equation*}
where $\eta$ denotes the design with masses $\tfrac{\overline{\omega}_1}{1-\overline{\omega}_{l+1}}, \ldots,\tfrac{\overline{\omega}_{l}}{1-\overline{\omega}_{l+1}}$ at the points $\overline{d}_1,\ldots,\overline{d}_l$, respectively. Therefore we obtain
\begin{equation*}
	(1-\overline{\omega}_{l+1}) M_1(  \eta,\theta_1) \geq (1-\tilde \omega_{k+1}) M_1(\tilde \xi,\theta_1) \geq (1-\overline{\omega}_{l+1}) M_1(  \xi,\theta_1).
\end{equation*}
Because the design $ \xi$ is admissible we have $M_1( \eta,\theta_1) = M_1( \xi,\theta_1)$. Using the block structure of the information matrix and the assumption that the design $\tilde \xi$ is not admissible it follows that
\begin{equation*}
	(\overline{\omega}_{l+1}-\tilde \omega_{k+1})M_1( \xi,\theta_1) \leq 0  \ \ \mbox{and} \ \ (\tilde \omega_{k+1}-\overline{\omega}_{l+1})\I(\theta_2) \leq 0.
\end{equation*}
This yields $\overline{\omega}_{l+1}=\tilde \omega_{k+1}$ and $M(\tilde \eta, \theta_1)=M(\tilde \xi, \theta_1)$, which is   a contradiction to the assumption that the design $\tilde \xi$ is not admissible. The desired result follows.
%%von Holger
%%\begin{equation}\label{proof1}
	%%M_1(\eta,\theta_1) \geq (1-\tilde\omega_{k+1}+\overline{\omega}_{l+1}) M_1(\xi,\theta_1) \geq M_1(\xi,\theta_1).
%%\end{equation}
%%Because $\xi$ is admissible it follows that $\eta=\xi$ and $\overline{\omega}_{l+1} \geq \tilde\omega_{l+1}$. However, in this case \eqref{proof1} implies $M_1(\eta,\theta)>M_1(\xi,\theta)$ which is a contradiction to the fact that $\eta=\xi$.
\hfill$\Box$

\begin{small}
 \setlength{\bibsep}{3pt}
\bibliographystyle{apalike}
\bibliography{drug}
\end{small}
\end{document}